\documentclass[a4paper]{cas-sc}

\usepackage[authoryear,longnamesfirst]{natbib}
\usepackage{amsmath}
\usepackage{amssymb}  
\usepackage{graphicx}

\usepackage{subcaption}
\usepackage{ulem}
\usepackage{diagbox}   
\usepackage{float}      
\usepackage{booktabs}   
\usepackage{array}      

\def\tsc#1{\csdef{#1}{\textsc{\lowercase{#1}}\xspace}}
\tsc{WGM}
\tsc{QE}
\tsc{EP}
\tsc{PMS}
\tsc{BEC}
\tsc{DE}

\begin{document}
\let\WriteBookmarks\relax
\def\floatpagepagefraction{1}
\def\textpagefraction{.001}

\shorttitle{Entropy-Based Vibration Data Quality Control}

\shortauthors{Yuan et~al.}

\title [mode = title]{Multi-Dimensional Entropy for Vibration Measurement Data Quality Assessment and Erroneous Signal Identification in Wind Turbines}



\author[1]{Xiao-Ming Yuan}
\cormark[1]
\orcidauthor{0009-0002-0977-585X}{Xiao-Ming Yuan}
\ead{yuanxiaoming@mywind.com.cn}
\ead{eeyxm@outlook.com}

\author[1]{Zishun Wang}
\ead{wangzishun@mywind.com.cn}

\author[1]{Donghui Zhao}
\ead{zhaodonghui@mywind.com.cn}

\author[1]{Deshui Li}
\cormark[2]
\ead{lideshui@mywind.com.cn}

\author[1]{Min Zhang}
\cormark[3]
\ead{zhangmin01@mywind.com.cn}

\affiliation[1]{organization={Mingyang Smart Energy Group Co., Ltd.},
    city={Zhong Shan},
    country={People's Republic of China}}

\cortext[1]{Corresponding author}
\cortext[2]{Corresponding author}
\cortext[3]{Corresponding author}


\begin{abstract}
Ensuring measurement data quality is essential for reliable condition monitoring of industrial wind turbine drivetrains, where vibration measurements can be affected by sensor malfunctions, turbine shutdown conditions, and other non-diagnostic states. Such invalid measurements may compromise the reliability of subsequent monitoring and data-driven analysis procedures. 
This study proposes a Multi-Dimensional Entropy (MDE) metric as a front-end data quality assessment and control mechanism for vibration measurement validity evaluation. By characterizing signal distributions from multiple perspectives, including time-domain amplitude, spectral amplitude, and frequency-band energy, MDE captures statistical differences between valid and erroneous vibration measurements. By integrating MDE and RMS as feature representations, lightweight machine learning models are employed as evaluation tools to assess the effectiveness of the proposed representation.
Experiments on a large-scale, heterogeneous real-world dataset—comprising 57,643 vibration samples collected from 12 wind farms and 14 turbine units, covering multiple drivetrain components, diverse sensor brands, and varying sampling configurations over long-term operation—demonstrate that the proposed method achieves over 99\% accuracy in identifying erroneous vibration measurements.
The proposed approach can be deployed as a front-end data quality gate before downstream signal processing, feature extraction, and condition monitoring procedures, ensuring that subsequent analyses are performed using reliable vibration measurements. The results demonstrate the robustness of MDE under heterogeneous sensor configurations and highlight its potential for industrial-scale vibration measurement quality assessment.
\end{abstract}



\begin{keywords}
Measurement Data Quality \sep  Data Quality Control \sep Wind Turbine Monitoring \sep Erroneous Signal Identification \sep Multi-Dimensional Entropy
\end{keywords}

\maketitle

\section{Introduction}

The proliferation of sensors in industrial systems generates vast amounts of data for automated condition monitoring. In wind turbine drivetrain monitoring, however, vibration signals are sometimes corrupted by sensor malfunctions and operational conditions such as turbine shutdowns. These issues give rise to erroneous vibration signals with abnormal amplitude distributions and distorted spectral characteristics, which fail to provide reliable information for subsequent monitoring and analysis. To address this issue, a reliable front-end data quality control mechanism is required to identify and filter such signals before downstream analysis. Without this step, erroneous inputs can significantly degrade downstream analysis, including condition monitoring, fault diagnosis, RUL estimation, and maintenance decision-making. Manual inspection by experienced engineers is labor-intensive and difficult to scale, especially for large-scale wind turbine fleets.

In this study, erroneous vibration signals are defined as measurements that fail to provide reliable information about the monitored systems. Such signals typically exhibit inconsistencies in their characteristic patterns across different domains, particularly in the time and frequency domains considered in this study. A visual comparison between normal and erroneous signals in both time and frequency domains is provided in Section~\ref{visual of erroneous signal}.

To address this problem, this paper proposes \textbf{Multi-Dimensional Entropy (MDE)}, a computationally efficient entropy-based metric for characterizing measurement validity from multiple signal representations. Instead of directly analyzing raw signals, MDE evaluates the distribution characteristics of signal patterns extracted from different domains, such as the time domain and the frequency domain. By integrating MDE and root mean square (RMS) as feature representations, lightweight machine learning models are employed as evaluation tools for validity identification. 

In this study, fault diagnosis is not performed; instead, the proposed method focuses on ensuring the reliability of vibration measurements before they are delivered to downstream monitoring algorithms.

The proposed method is validated using real-world wind turbine data collected from different wind farms, turbine types, and operational periods, covering multiple components and sensor configurations. Experimental results demonstrate that the method achieves high identification accuracy and strong robustness. In addition, the low computational complexity of the method enables seamless integration into existing monitoring and analysis frameworks.

\section{Related Work}

Vibration measurements are fundamental information sources for industrial condition monitoring systems. Reliable interpretation of vibration measurements requires not only effective feature extraction but also reliable input data quality. Fault diagnosis relies on vibration analysis, including wind turbines \cite{Randall2021}. Traditional fault diagnosis methods primarily rely on handcrafted features extracted from time and frequency domains (e.g., RMS, kurtosis, and sideband analysis), combined with empirical rules or shallow machine learning models for state identification \cite{Gao2015}. 

With the development of the field, fault diagnosis has evolved from experience-driven feature analysis to signal processing and statistical learning approaches, and more recently to data-driven and artificial intelligence-based paradigms \cite{Lei2020}. In particular, deep learning methods have been widely adopted for automatic feature extraction and end-to-end diagnosis \cite{Zhang2020}.


However, most existing studies are validated on publicly available datasets, which are typically limited in scale, collected under controlled conditions, and exhibit relatively high data quality. This creates a gap between current research and real industrial scenarios characterized by heterogeneous, noisy, and large-scale data.
In practical industrial environments, data quality plays a critical role in the performance of fault diagnosis systems. Low-quality or corrupted data can significantly degrade diagnostic accuracy and even lead to false or missed detections \cite{Chen2023}. As a result, data quality control has gradually become an important issue in condition monitoring systems. Existing studies have explored related topics such as abnormal data detection \cite{Wang2023, Zhang2022Access}, sensor fault diagnosis \cite{Li2017}, and data preprocessing techniques \cite{Famili1997}. In addition, data quality assurance has been investigated in other domains, such as environmental monitoring and meteorological measurements \cite{Foken2004, Steiner1999}. 
\textbf{Despite these efforts, most studies focus on improving the robustness of diagnostic models rather than explicitly assessing the validity of the input vibration signals themselves.} In particular, the problem of determining whether a vibration measurement is valid and suitable for subsequent analysis remains largely underexplored.

To better characterize the complexity and uncertainty of non-stationary vibration signals, entropy-based descriptors have been widely used to characterize complex measurement signals. Representative measures include Shannon entropy, permutation entropy, sample entropy, and their multiscale or multivariate extensions, which have demonstrated strong capability in capturing nonlinear dynamics and distributional characteristics \cite{Sabeti2009, Huo2020}. 
These entropy measures have been successfully applied in various contexts, including biomedical signal analysis and mechanical fault diagnosis, and have also been extended through approaches such as fuzzy entropy and multiscale symbolic entropy \cite{Xie2011, Li2022}. 
However, most existing entropy-based methods are confined to a single domain (e.g., time or frequency domain) and lack the ability to jointly model multi-domain information. Moreover, they are primarily designed for fault identification tasks, with limited attention paid to signal validity or data quality assessment.

In real-world industrial environments, vibration data are inherently heterogeneous and often contaminated by noise and disturbances. Existing studies have addressed challenges such as feature selection under redundant information \cite{Rauber2015}, noise-robust fault diagnosis \cite{He2025}, and representation learning using denoising autoencoders \cite{Jiang2018}. In addition, advanced signal processing and deep learning approaches have been proposed to enhance fault-related features under noisy and variable-speed conditions \cite{Guan2026, Wang2018}. 
Despite these efforts, most studies focus on extracting fault-related information from noisy signals, implicitly assuming that the input signals are valid. In practical wind turbine applications, however, vibration data are frequently affected by non-fault factors such as sensor malfunctions and turbine shutdown conditions. These erroneous signals typically exhibit abnormal amplitude distributions and distorted spectral patterns, yet do not contain any meaningful diagnostic information. 
The presence of such signals not only increases manual analysis effort but may also mislead automated diagnostic models, thereby degrading system reliability. Nevertheless, the problem of erroneous vibration signal identification has received limited attention, and no dedicated framework has been established for systematic data quality control in wind turbine monitoring systems.

To address this issue, this study proposes a Multi-Dimensional Entropy (MDE) framework for vibration measurement validity assessment and data quality control. The proposed method integrates entropy measures from multiple perspectives, including time-domain amplitude distribution, spectral amplitude distribution, and frequency-band energy distribution, and combines them with RMS to construct a compact feature representation. 
Furthermore, lightweight machine learning models are employed to enable efficient identification of erroneous signals. Compared with existing approaches, the proposed method provides a systematic framework for vibration measurement validity assessment and enables automatic data quality control before downstream monitoring and diagnostic procedures.
\section{Theoretical Background}
The proposed multi-dimensional entropy (MDE) method aims to characterize vibration signals from multiple dimensions, including both time and frequency domains. The core idea is to quantify variations in signal patterns through entropy-based measures. To this end, the vibration signal is first transformed into a probability density representation, upon which information entropy is calculated. The resulting entropy features are then used for subsequent classification. This section introduces the fundamental concepts required for the proposed method.

\subsection{Probability Density Representation of Vibration Signals}
\begin{figure}[pos=htbp]
  \centering
  \includegraphics[width=1\textwidth]{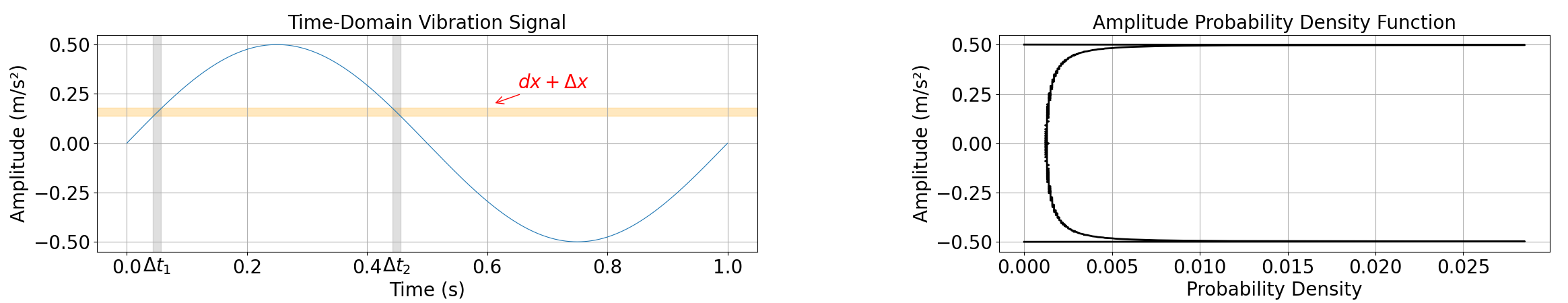}
  \caption{Left: time-domain waveform of a sine signal. Right: corresponding amplitude probability density function (PDF).}
  \label{fig:amplitude_probability_density_function}
\end{figure}

Prior to entropy calculation, the vibration signal is transformed into a probability density representation.Given a vibration signal, the probability density function (PDF) of its amplitude can be estimated. As illustrated in Fig.~\ref{fig:amplitude_probability_density_function}, a sine wave is presented as an example. The left panel shows the time-domain waveform, while the right panel illustrates the corresponding amplitude PDF. For any amplitude value \(x\), the PDF describes the likelihood of observing that amplitude in the signal.

Following standard vibration signal analysis practices \cite{Newland2005}, the PDF is  estimated using a histogram-based binning approach. The amplitude range of the signal is divided into a set of discrete bins, and the number of samples falling into each bin is counted to form a histogram. The histogram is then normalized by the total number of samples, resulting in a probability distribution whose integral equals unity. This representation transforms the raw vibration signal into a statistical description, enabling the characterization of amplitude distribution properties that are not directly observable in the time-domain waveform.

\subsection{Information Entropy for Signal Characterization}
Based on the probability density representation, information entropy can be used to quantify the distribution characteristics of vibration signals.
Information entropy, introduced by \citet{Shannon1948}, measures the uncertainty associated with a probability distribution. For a discrete signal \( x \) with \( N \) possible states and corresponding probabilities \( p(x_i) \), the entropy \( H \) is defined as:
\begin{equation}
    H = -\sum_{i=1}^{N} p(x_i) \log_2 p(x_i)
    \label{eq:info_entropy}
\end{equation}
where \( p(x_i) \) can be estimated using histogram-based methods for vibration signals.

For vibration data, information entropy reflects the dispersion and irregularity of signal amplitudes. Compared with normal signals, erroneous signals typically exhibit distinct distribution patterns in both amplitude range and frequency content. These differences lead to measurable variations in entropy values, which form the fundamental basis for erroneous signal identification in this study.

However, entropy derived from a single-domain representation is often insufficient to fully characterize the complex patterns observed in real-world vibration signals. In practice, differences between normal and erroneous signals may manifest simultaneously in time-domain morphology and frequency-domain structures. This limitation motivates the development of the proposed multi-dimensional entropy (MDE) framework, which extends entropy analysis across multiple complementary domains.

\subsection{Machine Learning Models for Validation Evaluation}

After extracting entropy-based features from vibration measurements, lightweight machine learning models are employed to evaluate the effectiveness of the proposed MDE representation for measurement validity identification. In this study, Support Vector Machine (SVM), Random Forest (RF), and LightGBM are selected as representative models due to their different learning mechanisms and computational efficiency.

SVM is a margin-based supervised learning method that constructs an optimal decision boundary by maximizing the separation margin between different classes \cite{Cortes1995}. RF is an ensemble learning approach that combines multiple decision trees to improve robustness and reduce overfitting \cite{Breiman2001}. LightGBM is a gradient boosting framework that sequentially builds decision trees to optimize prediction performance with high computational efficiency \cite{Ke2017}. 

In this study, these models are not considered as the main contribution but are used as evaluation tools to verify whether the proposed MDE representation can effectively distinguish valid and erroneous vibration measurements. The hyperparameters of all models are optimized through grid search, and detailed configurations are provided in Section~\ref{hyperparameters search}.

\section{Proposed Method}
To facilitate the development of the proposed MDE method for erroneous signal identification, this section first presents a comparative visualization of normal and erroneous vibration signals, and then introduces the formulation of the proposed MDE.

\subsection{Visualization of Erroneous Vibration Signals}
\label{visual of erroneous signal}
This subsection presents a comparative visualization of normal and erroneous vibration signals to illustrate their fundamental differences.

\begin{figure}[pos=htbp]
  \centering
  \begin{subfigure}[b]{0.48\textwidth}
    \centering
    \includegraphics[width=\textwidth]{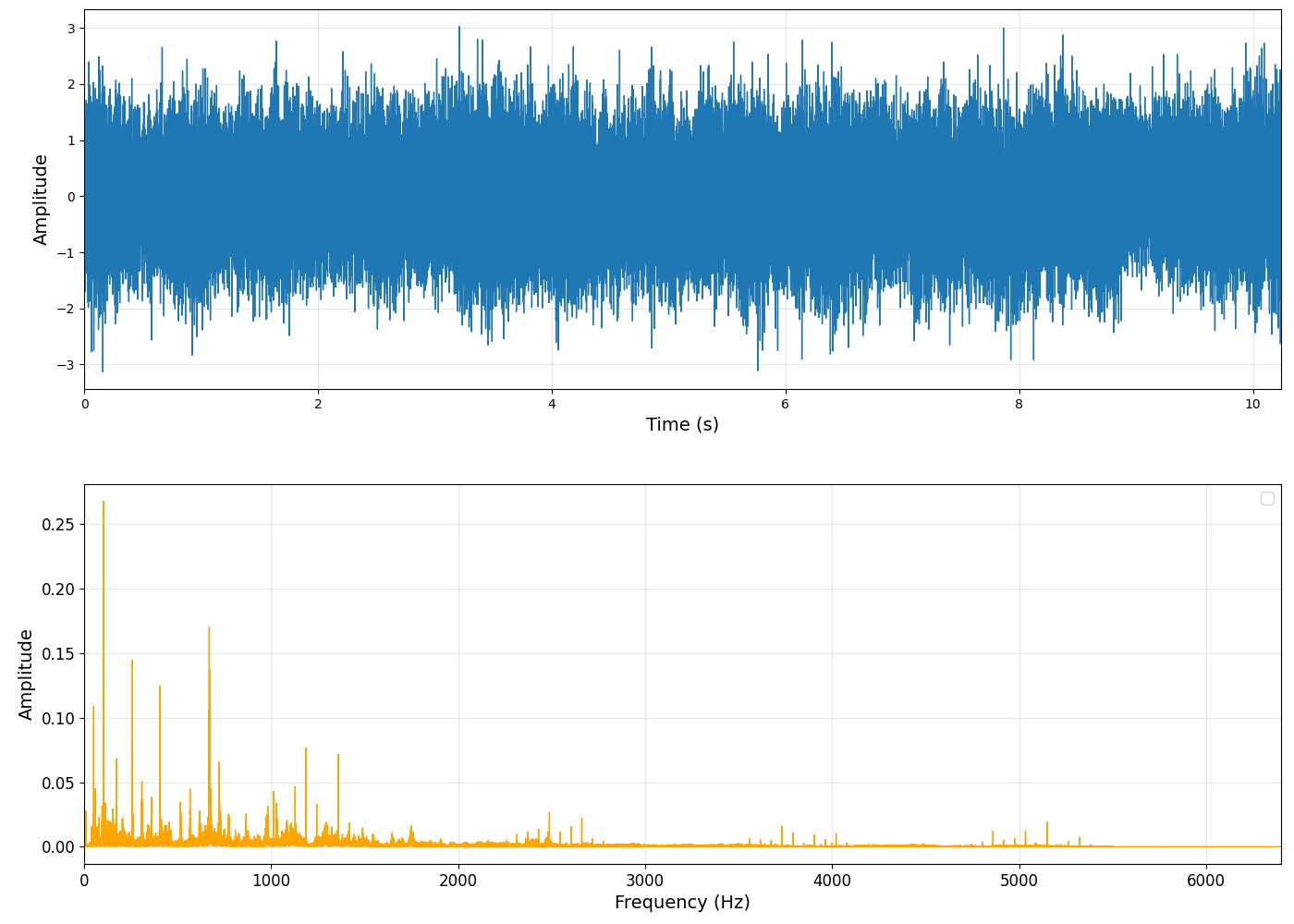}
    \caption{normal vibration signal 1}
    \label{fig:1a}
  \end{subfigure}
  \hfill
  \begin{subfigure}[b]{0.48\textwidth}
    \centering
    \includegraphics[width=\textwidth]{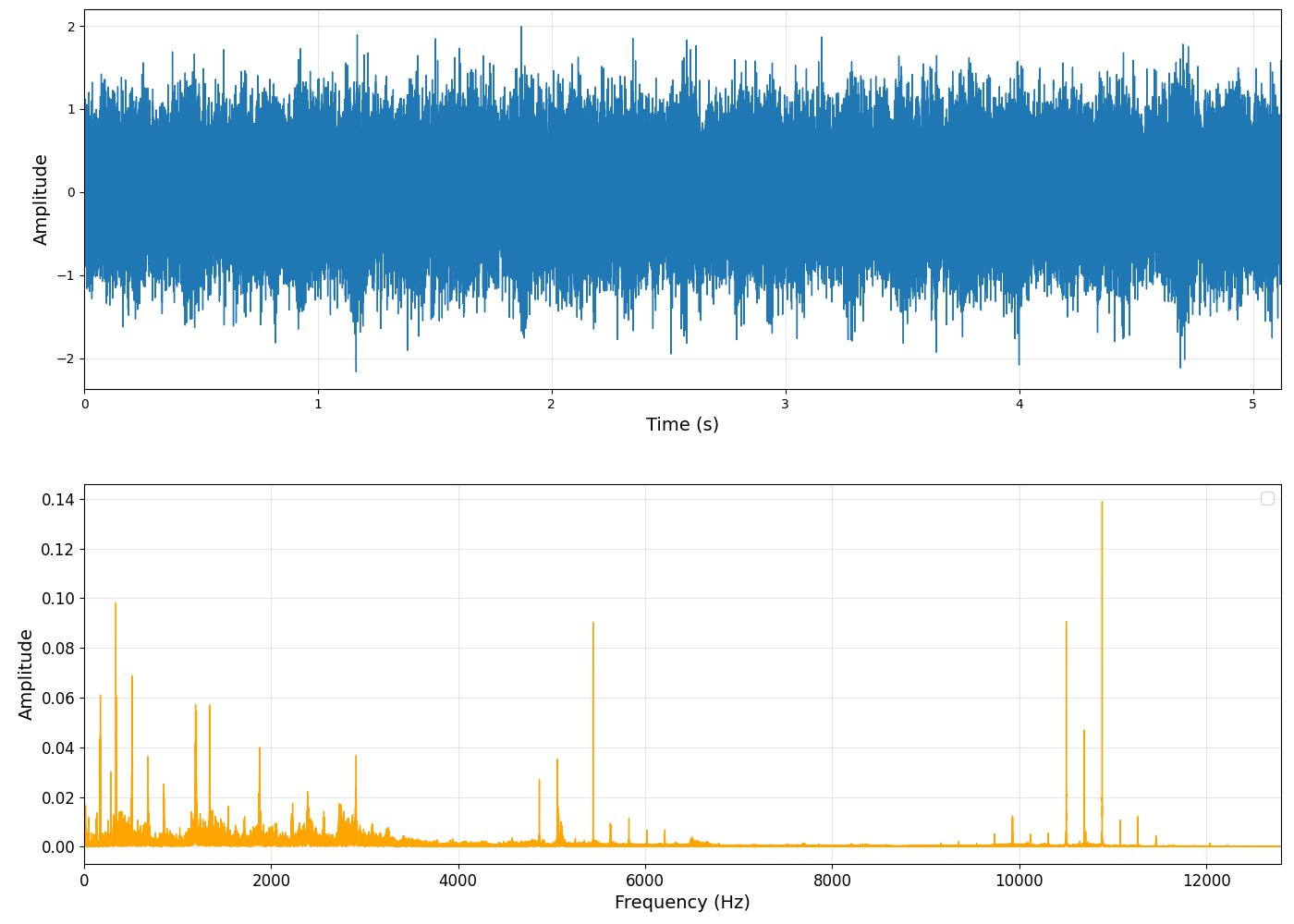}
    \caption{normal vibration signal 2}
    \label{fig:1b}
  \end{subfigure}
  
  \vskip\baselineskip
  \begin{subfigure}[b]{0.48\textwidth}
    \centering
    \includegraphics[width=\textwidth]{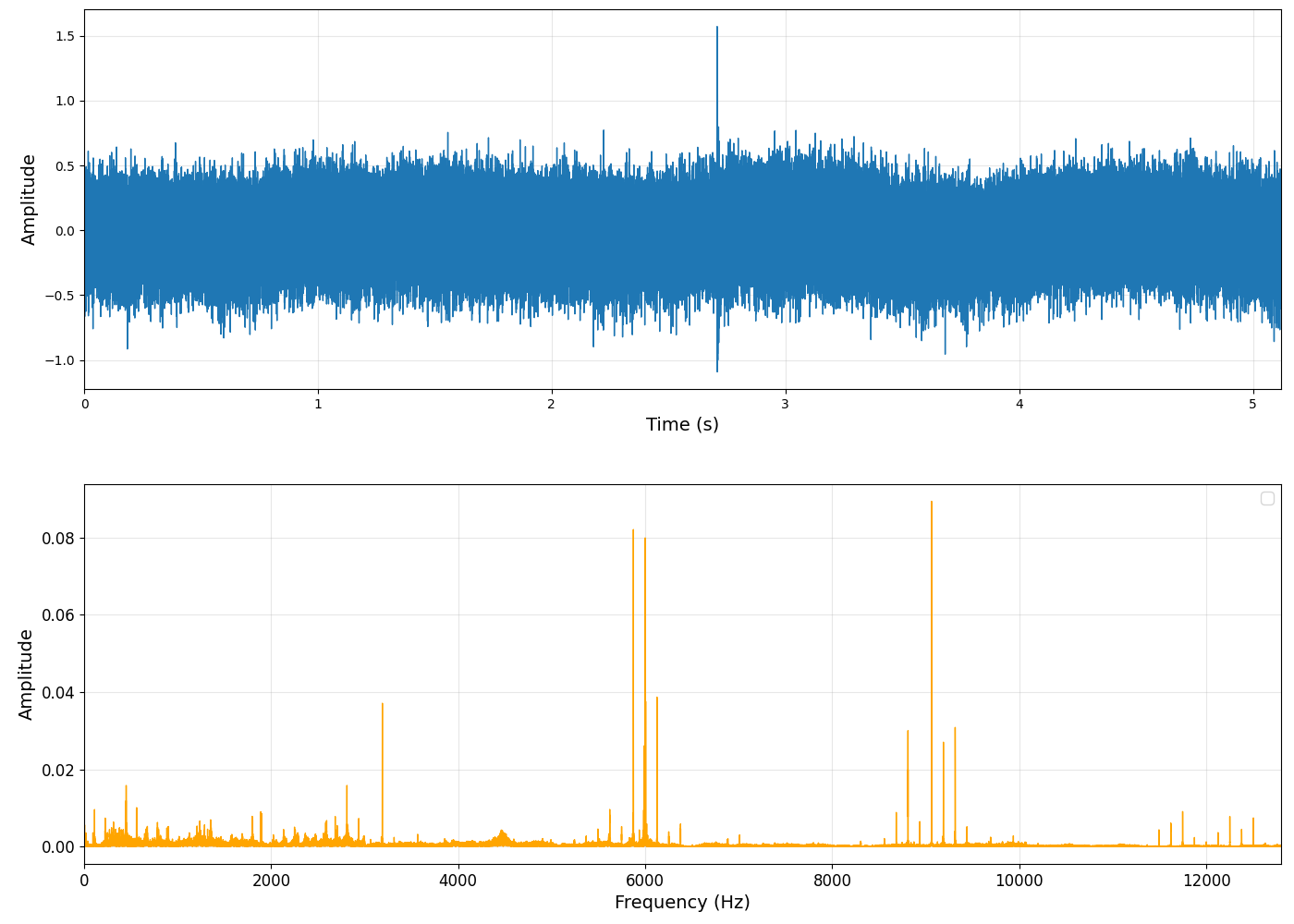}
    \caption{normal vibration signal 3}
    \label{fig:1c}
  \end{subfigure}
  \hfill
  \begin{subfigure}[b]{0.48\textwidth}
    \centering
    \includegraphics[width=\textwidth]{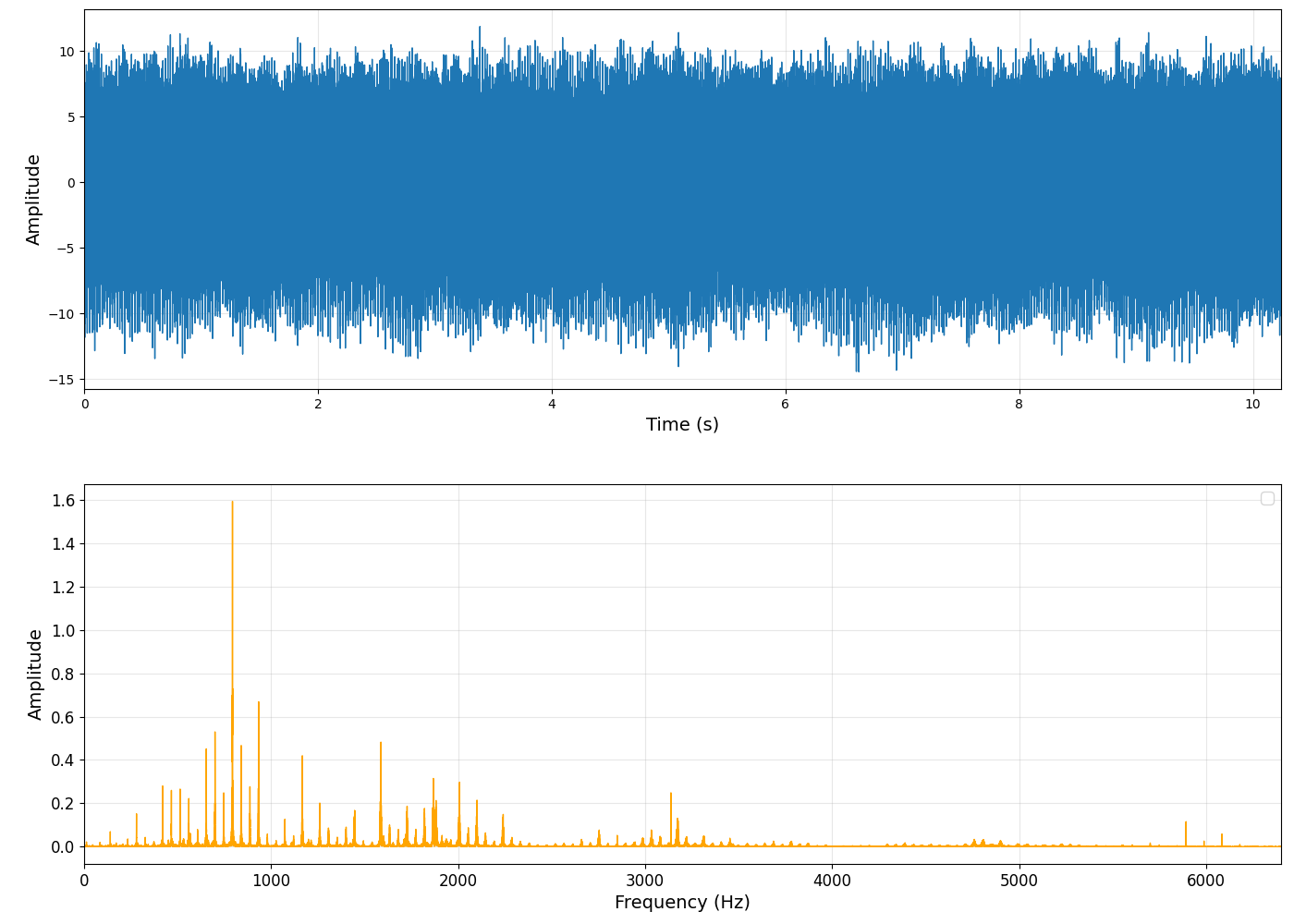}
    \caption{normal vibration signal 4}
    \label{fig:1d}
  \end{subfigure}

  \caption{Display of variant normal vibration acceleration signals.}
  \label{fig:normal_signal_display}
\end{figure}

The time-domain waveform and frequency spectrum of vibration acceleration signals contain essential information about the dynamic behavior and health condition of rotating mechanical components. In practical condition monitoring systems (CMS), characteristic frequencies and their sideband structures are commonly used to assess mechanical integrity. Fig.~\ref{fig:normal_signal_display} presents representative examples of normal vibration signals and erroneous vibration signals, illustrated in both time and frequency domains.

\begin{figure}[pos=htbp]
  \centering
  \begin{subfigure}[b]{0.48\textwidth}
    \centering
    \includegraphics[width=\textwidth]{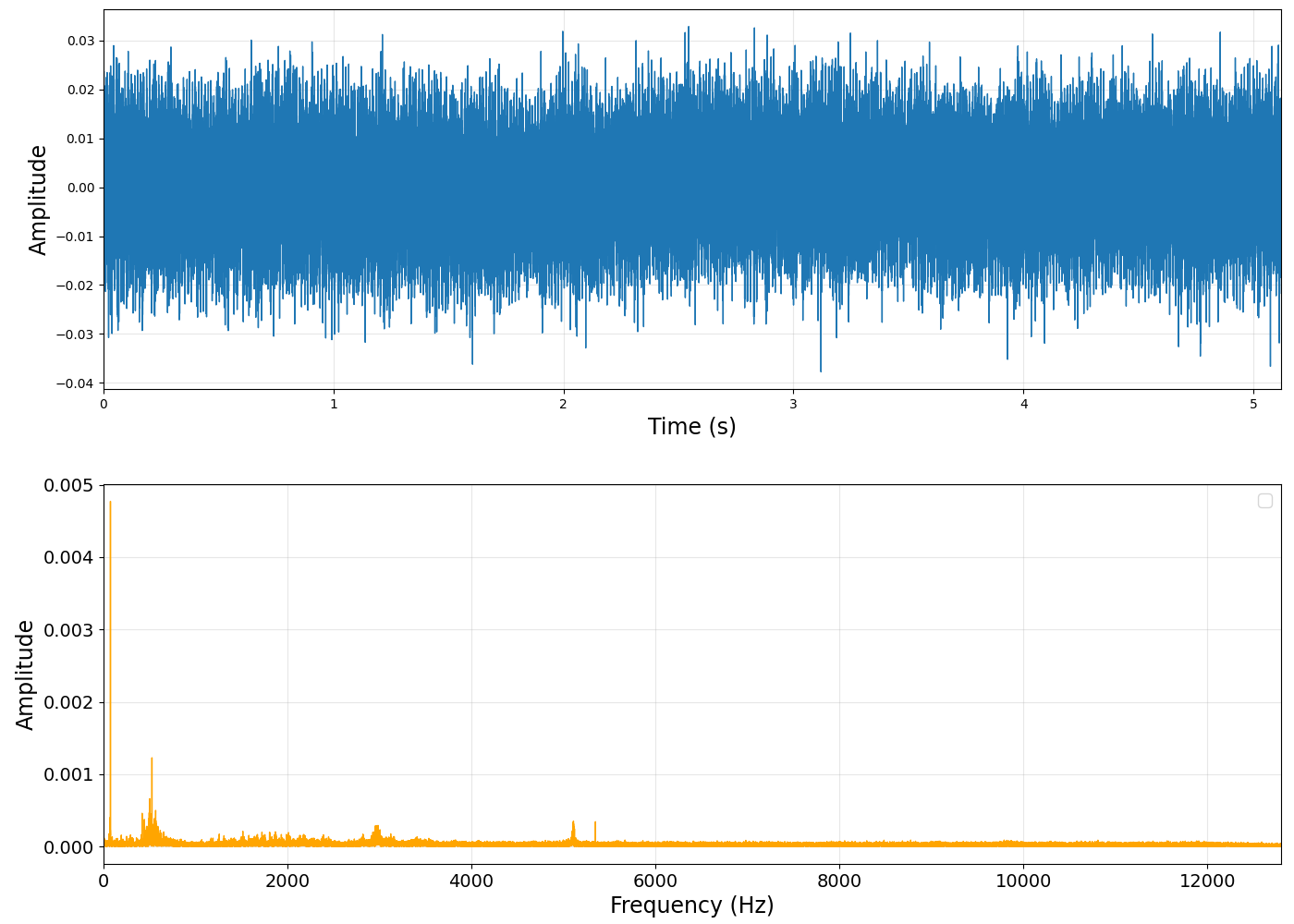}
    \caption{erroneous vibration signal 1}
    \label{fig:1a}
  \end{subfigure}
  \hfill
  \begin{subfigure}[b]{0.48\textwidth}
    \centering
    \includegraphics[width=\textwidth]{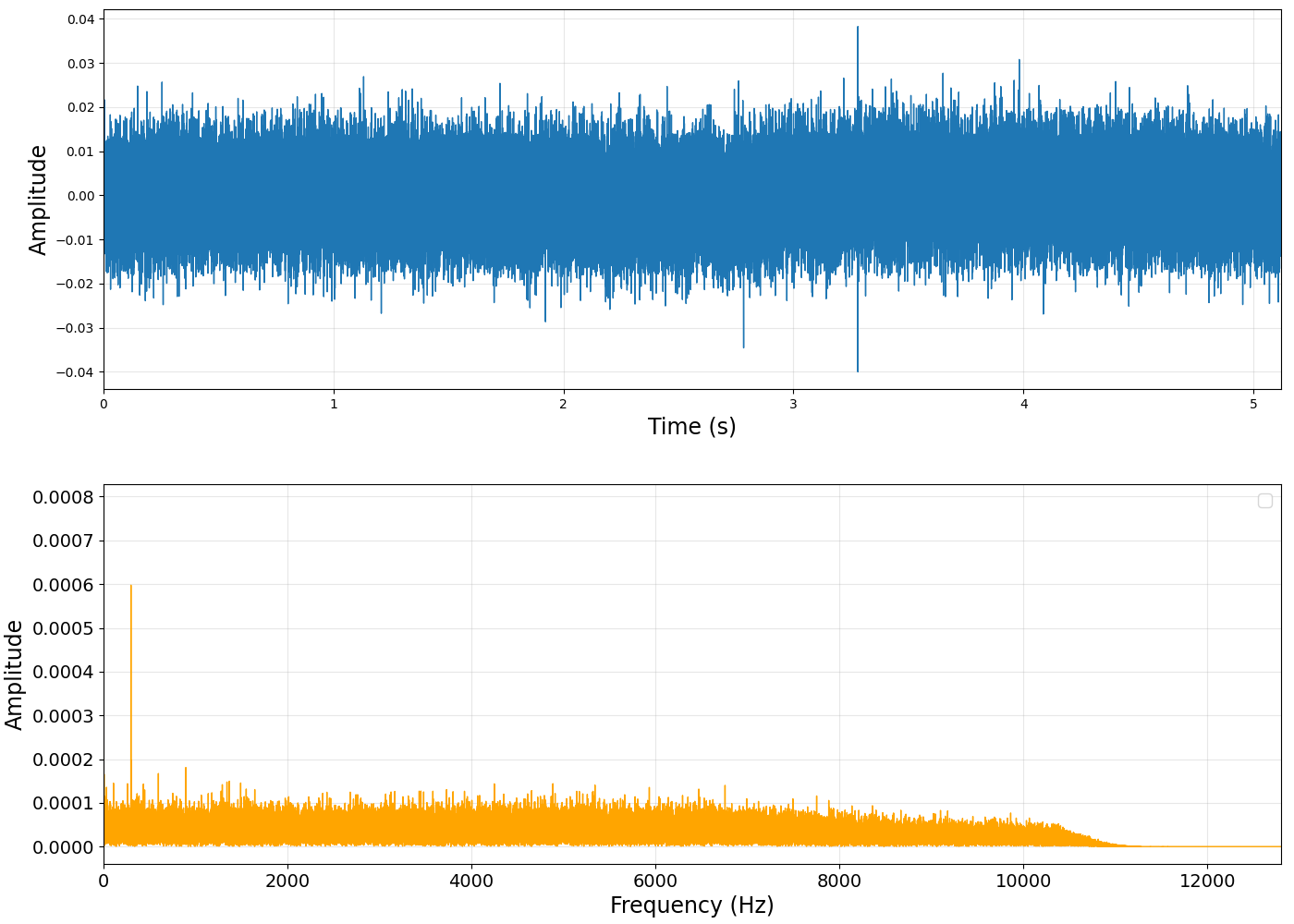}
    \caption{erroneous vibration signal 2}
    \label{fig:1b}
  \end{subfigure}
  
  \vskip\baselineskip
  \begin{subfigure}[b]{0.48\textwidth}
    \centering
    \includegraphics[width=\textwidth]{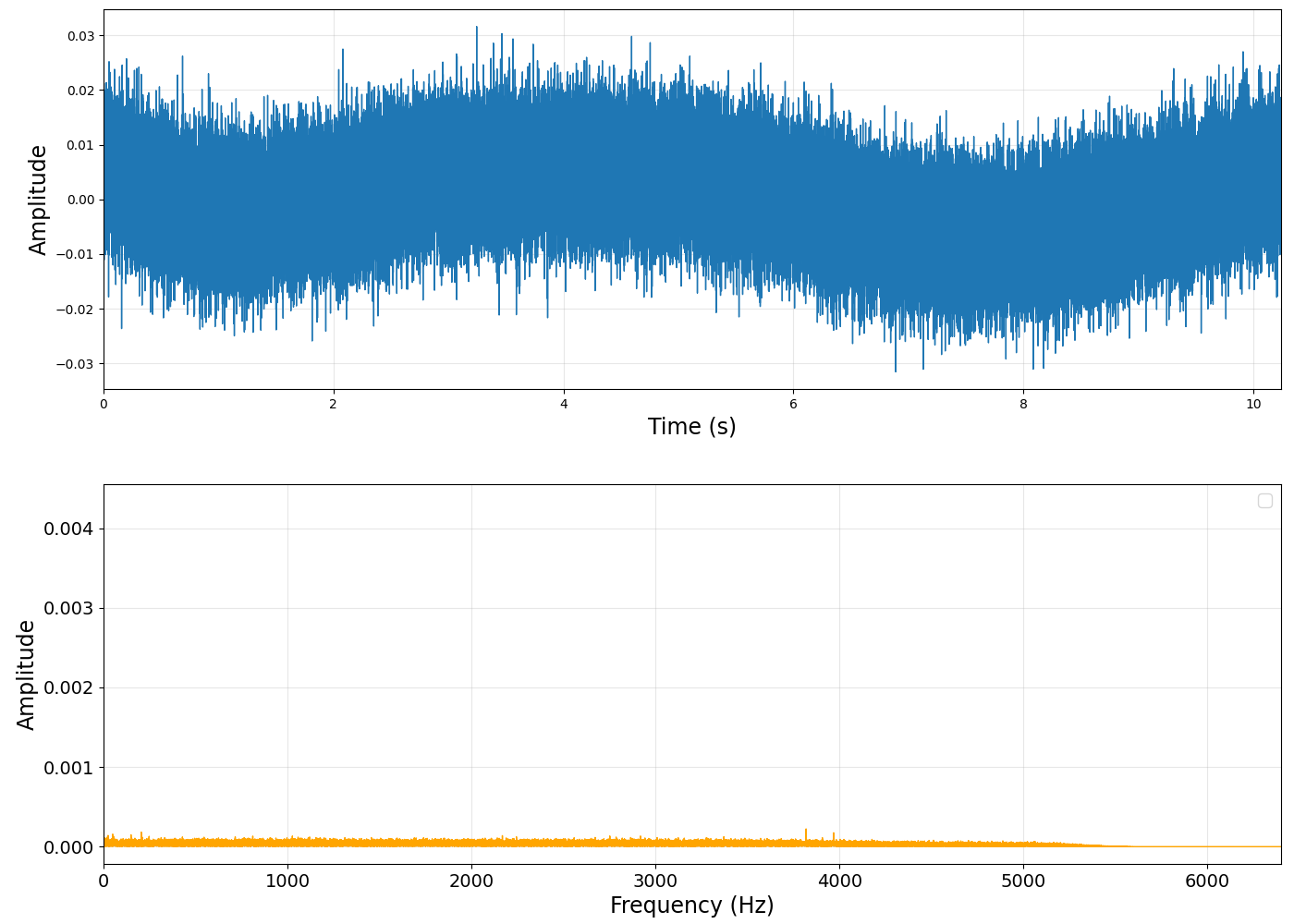}
    \caption{erroneous vibration signal 3}
    \label{fig:1c}
  \end{subfigure}
  \hfill
  \begin{subfigure}[b]{0.48\textwidth}
    \centering
    \includegraphics[width=\textwidth]{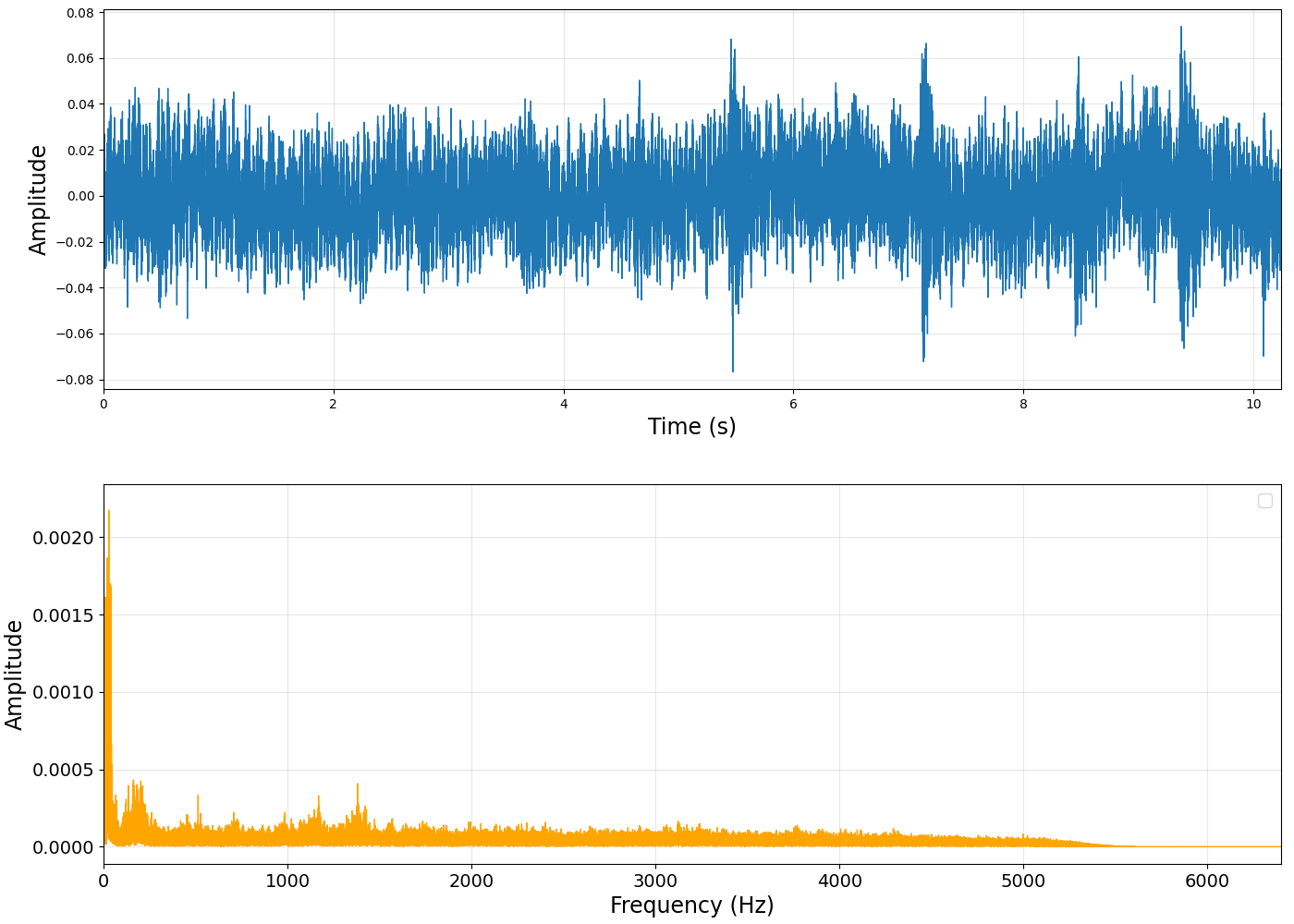}
    \caption{erroneous vibration signal 4}
    \label{fig:1d}
  \end{subfigure}

  \caption{Display of variant erroneous vibration acceleration signals.}
  \label{fig:false_signal_display}
\end{figure}

In contrast, Fig.~\ref{fig:false_signal_display} illustrates erroneous vibration signals, which fail to reflect the true dynamic response or mechanical condition of the rotating components. These signals are defined as \textbf{erroneous vibration signals}, as they contain no meaningful diagnostic information related to mechanical integrity.

This visual contrast highlights the fundamental difference between erroneous and normal vibration signals. From a statistical perspective, erroneous signals often exhibit concentrated distributions in either amplitude or frequency representations. In contrast, normal signals typically present richer and more structured patterns across both domains, reflecting underlying mechanical processes.

Such differences in distribution characteristics indicate that normal and erroneous vibration signals exhibit fundamentally different statistical patterns across multiple representations.

However, these differences are not always reliably distinguishable through direct observation or single-domain analysis, especially under real-world operating conditions.

Therefore, there is a need for robust quantitative indicators that can characterize signal variations in a consistent and unified manner. This requirement motivates the development of the proposed multi-dimensional entropy (MDE) framework.

\subsection{Multi-Dimensional Entropy}
\label{MDE formula}
To address the aforementioned requirement, this subsection introduces the proposed \textbf{Multi-Dimensional Entropy}.

The proposed MDE provides a unified information-theoretic framework for assessing vibration measurement validity by jointly considering multiple signal representations. Instead of directly analyzing raw signals, MDE maps the signal into corresponding probabilistic spaces by constructing probability density functions (PDFs) under different domain representations, and quantifies their statistical characteristics using entropy measures.

Within this framework, entropy is used to describe the distribution of signal-related quantities in different domains. In the time domain, entropy is associated with amplitude distribution. In the frequency domain, two complementary representations are considered: one based on spectral amplitudes, and the other based on energy distribution across frequency bands.

The entropy is computed from the corresponding PDF using a predefined binning scheme. This formulation enables a unified description of signal characteristics across multiple domains.

\subsubsection{Time-Domain Entropy (Signal Amplitude)}

In the time domain, the \textbf{Time-Amplitude Entropy (TAE)} quantifies the distribution of signal amplitudes. For a sinusoidal signal \(x(t)\), the calculation process is illustrated in Fig.~\ref{fig:amplitude_probability_density_function}. The figure consists of two panels: the left panel shows the original time-domain waveform, and the right panel displays the corresponding amplitude probability density function (PDF) obtained via histogram binning and normalization. The TAE is defined as:
\[
\mathrm{TAE} = -\sum_{i=1}^{N} p_1(x_i)\log_2 p_1(x_i),
\]
where \(N\) is the number of bins.

\begin{figure}[pos=htbp]
  \centering
  \includegraphics[width=\textwidth]{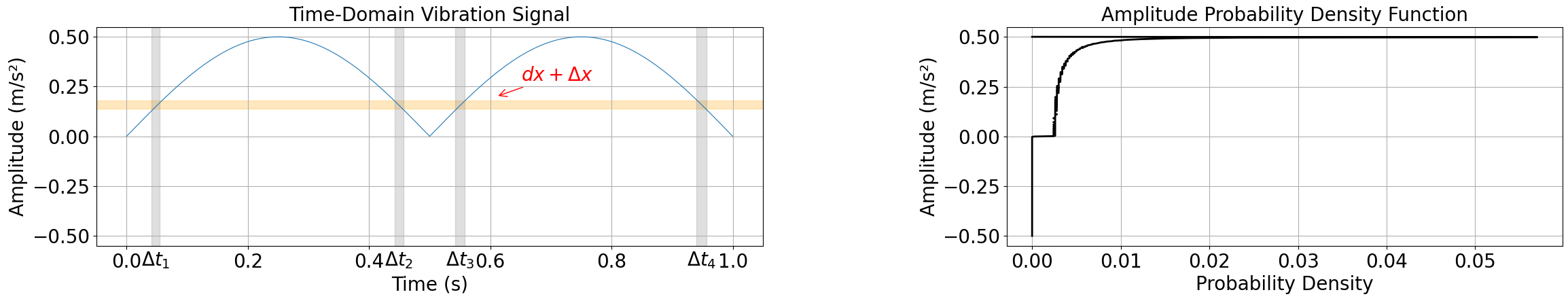}
  \caption{Time-domain waveform and corresponding amplitude probability density function (PDF) of the absolute signal.}
  \label{fig:absolute_amplitude_probability_density_function}
\end{figure}

A variant, the \textbf{Absolute Time-Amplitude Entropy (ATAE)}, is computed on the absolute-value signal \(|x(t)|\), as illustrated in Fig.~\ref{fig:absolute_amplitude_probability_density_function}. The figure consists of two panels: the left panel shows the time-domain waveform of the absolute signal, and the right panel displays the corresponding amplitude probability density function (PDF) obtained using the same binning and normalization procedure. The ATAE is given by:
\[
\mathrm{ATAE} = -\sum_{i=1}^{N} p_2(x_i)\log_2 p_2(x_i),
\]
This formulation increases sensitivity to amplitude concentration.

\subsubsection{Frequency-Domain Entropy (Spectrum Amplitude)}
In the frequency domain, the \textbf{Spectrum-Amplitude Entropy (SpAE)} characterizes the distribution of spectral amplitudes. 

The frequency spectrum of \(x(t)\) is obtained through the Fast Fourier Transform (FFT), retaining the single-sided spectrum scaled by \(2/n\) (where \(n\) is the signal length). The PDF of spectral amplitudes, \(p_3(x_i)\), is estimated by binning and normalization. The SpAE is expressed as:
\[
\mathrm{SpAE} = -\sum_{i=1}^{N} p_3(x_i)\log_2 p_3(x_i),
\]
where \(N\) is the total number of bins. 

\begin{figure}[pos=htbp]
  \centering
  \includegraphics[width=\textwidth]{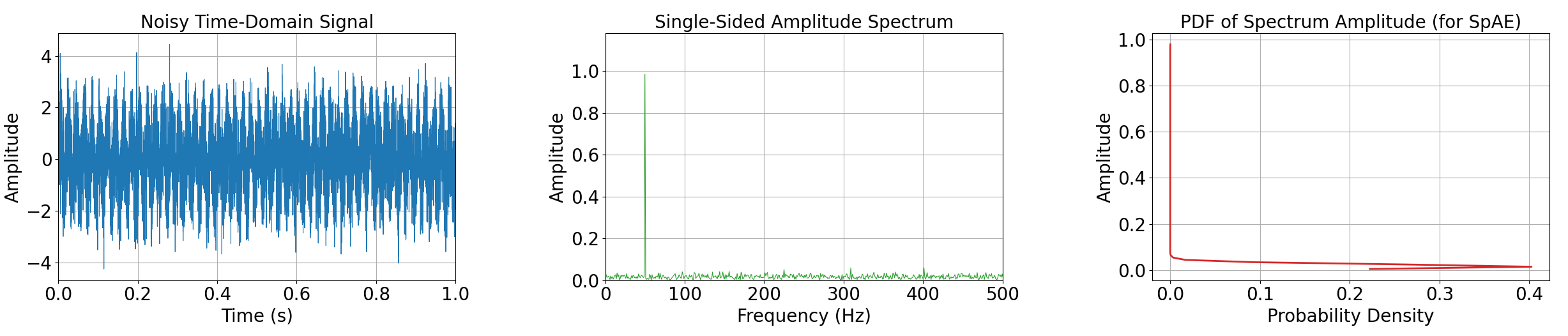}
  \caption{Illustration of the Spectrum-Amplitude Entropy (SpAE) calculation, including FFT transformation, single-sided scaling, and PDF estimation.}
  \label{fig:SpAE_process}
\end{figure}

The complete SpAE calculation process is demonstrated in Fig.~\ref{fig:SpAE_process} using a noisy sinusoidal signal \(x(t) = \sin(2\pi 50 t) + \eta(t)\), where \(\eta(t)\) denotes additive noise. The figure consists of three panels: the left panel shows the original time-domain waveform, the middle panel presents the the single-sided power spectrum is obtained via the FFT, scaled by \(2/n\) (where \(n\) is the signal length), which with a dominant peak at 50~Hz and broadband low-amplitude noise, and the right panel displays the corresponding PDF computed with a bin width of 0.01. Most spectral amplitudes are confined to the low-amplitude region, consistent with a concentrated energy pattern.

\subsubsection{Frequency-Band Energy Entropy}

To characterize the distribution of spectral energy across frequency bands, the \textbf{Frequency-Band Energy Entropy (FBEE)} is introduced. As illustrated in Fig.~\ref{fig:normal_signal_display}, normal vibration signals exhibit amplitudes along the frequency axis, indicating energy concentration in characteristic bands. In contrast, erroneous vibration signals  in Fig.~\ref{fig:false_signal_display} display nearly uniform energy across most frequency bands, reflecting some particular patterns comparing the normal signal.

\begin{figure}[pos=htbp]
  \centering
  \includegraphics[width=\textwidth]{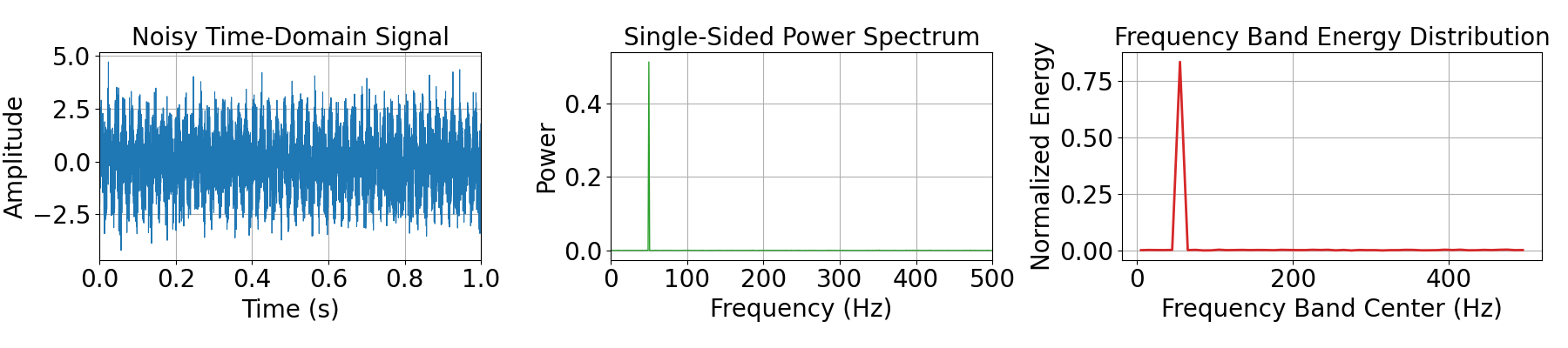}
  \caption{Comparison between normal and erroneous vibration signal spectra, highlighting the difference in frequency-band energy dispersion.}
  \label{fig:FBEE}
\end{figure}

The FBEE computation process for a noisy sinusoidal signal \(x(t) = \sin(2\pi 50 t) + \eta(t)\) (where \(\eta(t)\) denotes additive noise) is shown in Fig.~\ref{fig:FBEE}. The figure consists of three panels: the left panel shows the original time-domain waveform, the middle panel presents the single-sided power spectrum, and the right panel displays the corresponding PDF of the normalized band energies. First, the single-sided power spectrum is obtained via the Fast Fourier Transform (FFT), scaled by \(2/n\) (where \(n\) is the signal length) and squared. The spectrum is then partitioned into 10~Hz frequency bands, and the energy in each band \(E_i\) is normalized to form a probability distribution:
\[
p_4(x_i) = \frac{E_i}{\sum_{j=1}^{N} E_j}.
\]
The Frequency-Band Energy Entropy is subsequently calculated from this distribution:
\[
\mathrm{FBEE} = -\sum_{i=1}^{N} p_4(x_i)\log_2 p_4(x_i).
\]

In practical monitoring, sensors operate at diverse sampling frequencies (e.g., 2560~Hz, 5120~Hz, 12800~Hz, and 25600~Hz). To ensure cross-sensor comparability, the frequency range is standardized to the maximum Nyquist frequency (e.g., 12800~Hz for 25600~Hz sampling). For signals with lower sampling rates, zero-padding is applied to the unmeasured high-frequency range. Since bins with zero energy contribute nothing to entropy (\(0\log_2 0 = 0\)), this normalization ensures consistent FBEE values across different sampling rates.

Consider two equivalent signals: \(x_{\text{low}}(t)\) sampled at 12800~Hz and \(x_{\text{high}}(t)\) at 25600~Hz, both representing the same physical process below 6400~Hz. The corresponding entropy values remain identical because the additional zero-energy bins in the higher sampling case contribute no extra information. Thus, the FBEE maintains invariance with respect to sampling frequency, while retaining its ability to distinguish erroneous vibration signals—characterized by uniformly distributed band energies—from valid operational data with concentrated energy in characteristic frequency regions.

\subsection{Robustness Verification of MDE under Different Measurement Configurations}
\label{Robustness of MDE under different sapling config}
To verify the stability of the proposed Multi-Dimensional Entropy (MDE) framework with respect to sampling frequency and signal duration, a controlled simulation study was conducted using synthetically generated vibration signals. The objective is to assess whether MDE remain invariant when the same physical process is sampled at different rates and durations, provided that the total number of sampling points is identical.

In practical applications, vibration sensors operate at various sampling rates (e.g., 2.56~kHz, 5.12~kHz, 12.8~kHz, and 25.6~kHz). In this study, two representative configurations are considered: 12.8~kHz and 25.6~kHz. Two equivalent signals are constructed to represent the same underlying vibration process:
\[
x_{\text{low}}(t): f_s = 12800~\text{Hz}, \ \text{duration} = 10.24~\text{s}, \quad
x_{\text{high}}(t): f_s = 25600~\text{Hz}, \ \text{duration} = 5.12~\text{s},
\]
such that both have the same total number of samples ($N = 131072$). This ensures that any difference in entropy arises from sampling configuration rather than data length.

Four types of synthetic vibration signals were designed for the experiment:
\begin{enumerate}[(1)]
\item $x_1(t)$ — low-frequency components only;
\item $x_2(t)$ — both low- and high-frequency components;
\item $x_3(t)$ — $x_1(t)$ with added Gaussian noise;
\item $x_4(t)$ — $x_2(t)$ with added Gaussian noise.
\end{enumerate}

The signal components are defined as:
\[
\begin{aligned}
x_{\text{lower}}(t) &= \sin(2\pi \times 50t) + 0.5\sin(2\pi \times 150t) + 0.3\sin(2\pi \times 250t) + 0.8\sin(2\pi \times 750t) \\
&\quad + 0.1\sin(2\pi \times 1050t) + 1.5\sin(2\pi \times 2550t) + 1.8\sin(2\pi \times 3250t) + 3.3\sin(2\pi \times 4250t),\\
x_{\text{higher}}(t) &= 0.3\sin(2\pi \times 6250t) + 0.1\sin(2\pi \times 8250t) + 0.2\sin(2\pi \times 9250t) + 0.1\sin(2\pi \times 12500t),\\
\eta(t) &= k \times \mathcal{N}(0,1),
\end{aligned}
\]
where \(k\) controls the noise intensity.


The four signal groups are then defined as:
\[
\begin{aligned}
x_1(t) &= x_{\text{lower}}(t), \\
x_2(t) &= x_{\text{lower}}(t) + x_{\text{higher}}(t), \\
x_3(t) &= x_{\text{lower}}(t) + \eta(t), \\
x_4(t) &= x_{\text{lower}}(t) + x_{\text{higher}}(t) + \eta(t).
\end{aligned}
\]

For each signal group, two versions are generated:
\[
\{x_{i,\text{low}}(t),\,x_{i,\text{high}}(t)\}, \quad i = 1,2,3,4,
\]
corresponding to the two sampling configurations. The four entropy indicators—Time-Amplitude Entropy (TAE), Absolute Time-Amplitude Entropy (ATAE), Spectrum-Amplitude Entropy (SpAE), and Frequency-Band Energy Entropy (FBEE)—are computed and compared.

For the noise-included cases ($x_3$ and $x_4$), the results are evaluated under stochastic perturbations to assess statistical robustness. The results presented in Tables~\ref{tab:MDE_groups_k=0.2}, \ref{tab:MDE_groups_k=0.35}, \ref{tab:MDE_groups_k=0.5}, and \ref{tab:MDE_groups_k=1} demonstrate that the proposed entropy indicators exhibit minimal sensitivity to sampling rate and signal duration under controlled conditions. This observation is further supported by the results on real industrial data (Table~\ref{tab:classifier-performance} and Appendix~\ref{app:hyperparameters}), where mixed sampling configurations still yield consistently high identification performance.

\begin{table}[H]
\centering
\scriptsize
\renewcommand{\arraystretch}{1.1}
\caption{Comparison of four signal groups (Group 1--4) with k=0.2 random noise under different sampling configurations. Each subtable lists entropy indices (TAE, ATAE, SpAE, FBEE) computed at low and high sampling frequencies, with absolute differences $|\Delta_i|$}
\label{tab:MDE_groups_k=0.2}
\begin{tabular}{
  >{\centering\arraybackslash}p{0.45\textwidth}
  >{\centering\arraybackslash}p{0.45\textwidth}
}
\begin{tabular}{|c|c|c|c|c|}
\hline
\diagbox{Signal}{Index} & TAE & ATAE & SpAE & FBEE \\
\hline
$x_{1,\text{low}}$  & 6.91406 & 5.85938 & 0.00213 & 1.70990 \\
\hline
$x_{1,\text{high}}$ & 7.84471 & 6.81441 & 0.00213 & 1.70990 \\
\hline
$|\Delta_1|/\%$ & 0.97065 & 0.95503 & 0.0 & 0.0 \\
\hline
$|\Delta_1|/x_{1,low}\%$ & $14.039\%$ & $16.299\%$ & 0.0 & 0.0 \\
\hline
\end{tabular}
&
\begin{tabular}{|c|c|c|c|c|}
\hline
\diagbox{Signal}{Index} & TAE & ATAE & SpAE & FBEE \\
\hline
$x_{2,\text{low}}$  & 7.87986& 6.86685 & 0.00309 & 1.77128 \\
\hline
$x_{2,\text{high}}$ & 8.63812 & 7.65086 & 0.00309 & 1.77588 \\
\hline
$|\Delta_2|/\%$ & 0.75826 & 0.78401 & 0.0 & 0.0046 \\
\hline
$|\Delta_2|/x_{2,low}\%$ & $9.623\%$ & $11.417\%$ & 0.0 & $2.597\%$ \\
\hline
\end{tabular} 
\\
\textbf{Group 1} & \textbf{Group 2} \\[6pt]

\begin{tabular}{|c|c|c|c|c|}
\hline
\diagbox{Signal}{Index} & TAE & ATAE & SpAE & FBEE \\
\hline
$x_{3,\text{low}}$  & 10.18814 & 9.19191  &0.00213  & 1.74053 \\
\hline
$x_{3,\text{high}}$ & 10.23081 & 9.23521 & 0.00213  & 1.75304 \\
\hline
$|\Delta_3|/\%$ & 0.04267 & 0.04330 & 0.0 & 0.01251 \\
\hline
$|\Delta_3|/x_{3,low}\%$ & $0.419\%$ & $0.471\%$ &0.0 & $0.719\%$ \\
\hline
\end{tabular}
&
\begin{tabular}{|c|c|c|c|c|}
\hline
\diagbox{Signal}{Index} & TAE & ATAE & SpAE & FBEE \\
\hline
$x_{4,\text{low}}$  & 10.20022 & 9.20437 & 0.00313 & 1.79785 \\
\hline
$x_{4,\text{high}}$ & 10.23779 & 9.24215 & 0.00316 & 1.81593 \\
\hline
$|\Delta_4|$ & 0.03757 & 0.03778 & 0.00003 & 0.01808 \\
\hline
$|\Delta_4|/x_{4,low}\%$ & $0.368\%$ & $0.410\%$ & $0.958\%$ & $1.006\%$ \\
\hline
\end{tabular}
\\
\textbf{Group 3} & \textbf{Group 4} \\
\end{tabular}

\end{table}

\begin{table}[H]
\centering
\scriptsize
\renewcommand{\arraystretch}{1.1}
\caption{Comparison of four signal groups (Group 1--4) with k=0.35 random noise under different sampling configurations. Each subtable lists entropy indices (TAE, ATAE, SpAE, FBEE) computed at low and high sampling frequencies, with absolute differences $|\Delta_i|$}
\label{tab:MDE_groups_k=0.35}
\begin{tabular}{
  >{\centering\arraybackslash}p{0.45\textwidth}
  >{\centering\arraybackslash}p{0.45\textwidth}
}

\begin{tabular}{|c|c|c|c|c|}
\hline
\diagbox{Signal}{Index} & TAE & ATAE & SpAE & FBEE \\
\hline
$x_{3,\text{low}}$  & 10.23800  & 9.24252 & 0.00213  & 1.79166 \\
\hline
$x_{3,\text{high}}$ & 10.25930 & 9.26363 & 0.00213  & 1.82494 \\
\hline
$|\Delta_3|/\%$ & 0.02130 & 0.02111 & 0.0 & 0.03328 \\
\hline
$|\Delta_3|/x_{3,low}\%$ & $0.208\%$ & $0.228\%$ &0.0 & $1.857\%$ \\
\hline
\end{tabular}
&
\begin{tabular}{|c|c|c|c|c|}
\hline
\diagbox{Signal}{Index} & TAE & ATAE & SpAE & FBEE \\
\hline
$x_{4,\text{low}}$  & 10.24360  & 9.24834 & 0.00313 & 1.84438 \\
\hline
$x_{4,\text{high}}$ & 10.26491 & 9.26939 & 0.00316 & 1.88381  \\
\hline
$|\Delta_4|$ & 0.02131 & 0.02105 & 0.00003 & 0.03943 \\
\hline
$|\Delta_4|/x_{4,low}\%$ & $0.208\%$ & $0.228\%$ & $0.958\%$ & $2.138\%$ \\
\hline
\end{tabular}
\\
\textbf{Group 3} & \textbf{Group 4} \\
\end{tabular}

\end{table}

\begin{table}[H]
\centering
\scriptsize
\renewcommand{\arraystretch}{1.1}
\caption{Comparison of four signal groups (Group 1--4) with k=0.5 random noise under different sampling configurations. Each subtable lists entropy indices (TAE, ATAE, SpAE, FBEE) computed at low and high sampling frequencies, with absolute differences $|\Delta_i|$}
\label{tab:MDE_groups_k=0.5}
\begin{tabular}{
  >{\centering\arraybackslash}p{0.45\textwidth}
  >{\centering\arraybackslash}p{0.45\textwidth}
}

\begin{tabular}{|c|c|c|c|c|}
\hline
\diagbox{Signal}{Index} & TAE & ATAE & SpAE & FBEE \\
\hline
$x_{3,\text{low}}$  & 10.26552   & 9.27008 & 0.00213  & 1.85955 \\
\hline
$x_{3,\text{high}}$ & 10.27859 & 9.28321 & 0.00213  & 1.92075\\
\hline
$|\Delta_3|/\%$ & 0.01307 & 0.01313 & 0.0 & 0.06120 \\
\hline
$|\Delta_3|/x_{3,low}\%$ & $0.127\%$ & $0.142\%$ &0.0 & $3.291\%$ \\
\hline
\end{tabular}
&
\begin{tabular}{|c|c|c|c|c|}
\hline
\diagbox{Signal}{Index} & TAE & ATAE & SpAE & FBEE \\
\hline
$x_{4,\text{low}}$  & 10.27344  & 9.27827 & 0.00313 & 1.90757 \\
\hline
$x_{4,\text{high}}$ & 10.28275 & 9.28761 & 0.00316 &  1.97524  \\
\hline
$|\Delta_4|$ & 0.00931 & 0.00934 & 0.00003 & 0.06767 \\
\hline
$|\Delta_4|/x_{4,low}\%$ & $0.0906\%$ & $0.101\%$ & $0.958\%$ & $3.547\%$ \\
\hline
\end{tabular}
\\
\textbf{Group 3} & \textbf{Group 4} \\
\end{tabular}

\end{table}

\begin{table}[H]
\centering
\scriptsize
\renewcommand{\arraystretch}{1.1}
\caption{Comparison of four signal groups (Group 1--4) with k=1 random noise under different sampling configurations. Each subtable lists entropy indices (TAE, ATAE, SpAE, FBEE) computed at low and high sampling frequencies, with absolute differences $|\Delta_i|$}
\label{tab:MDE_groups_k=1}
\begin{tabular}{
  >{\centering\arraybackslash}p{0.45\textwidth}
  >{\centering\arraybackslash}p{0.45\textwidth}
}

\begin{tabular}{|c|c|c|c|c|}
\hline
\diagbox{Signal}{Index} & SAE & ASAE & SpAE & FBEE \\
\hline
$x_{3,\text{low}}$  & 10.34599 & 9.35124 & 0.23400 & 2.15488 \\
\hline
$x_{3,\text{high}}$ & 10.34573 & 9.35055 & 0.23057  & 2.34203 \\
\hline
$|\Delta_3|/\%$ & 0.00026 & 0.00069 & 0.0034 & 0.18715 \\
\hline
$|\Delta_3|/x_{3,low}\%$ & $0.0025\%$ & $0.0074\%$ & $1.466\%$ & $8.68\%$ \\
\hline
\end{tabular}
&
\begin{tabular}{|c|c|c|c|c|}
\hline
\diagbox{Signal}{Index} & SAE & ASAE & SpAE & FBEE \\
\hline
$x_{4,\text{low}}$  & 10.35282  & 9.35826 & 0.23265 & 2.18871 \\
\hline
$x_{4,\text{high}}$ & 10.34834 & 9.35416 & 0.22872 &  2.38235 \\
\hline
$|\Delta_4|$ & 0.00448 & 0.0041 & 0.00393 & 0.19364 \\
\hline
$|\Delta_4|/x_{4,low}\%$ & $0.0433\%$ & $0.0438\%$ & $1.689\%$ & $8.847\%$ \\
\hline
\end{tabular}
\\
\textbf{Group 3} & \textbf{Group 4} \\
\end{tabular}

\end{table}

From the results in Tables~\ref{tab:MDE_groups_k=0.2}, \ref{tab:MDE_groups_k=0.35}, \ref{tab:MDE_groups_k=0.5}, and \ref{tab:MDE_groups_k=1}, several consistent observations can be drawn. First, for all signal groups and noise levels, the entropy values computed under different sampling configurations remain highly consistent. In particular, SpAE exhibits near-identical values across all cases, while TAE and ATAE show only minor deviations, indicating that the proposed entropy measures are inherently insensitive to changes in sampling frequency and signal duration when the total number of samples is preserved. Second, as the noise level increases (from $k=0.2$ to $k=1$), the absolute differences between low- and high-sampling configurations generally decrease for TAE and ATAE, suggesting that stochastic perturbations tend to homogenize the statistical distributions and improve the consistency of entropy estimation. Third, FBEE shows slightly larger deviations compared to the other indicators, especially at higher noise levels. Nevertheless, the relative differences remain within an acceptable range. 

Overall, these results verify that the proposed MDE framework maintains stable and consistent behavior across varying sampling configurations and noise conditions, providing a solid foundation for its application to real-world datasets with heterogeneous sensor settings.

\subsection{Practical Considerations in Entropy-Based Feature Extraction}
In practical vibration monitoring, maintaining unit consistency in acceleration signals is essential. Commercial sensors typically output measurements in either \textbf{g} or \textbf{m/s$^2$}, where \(1\,\mathrm{g} \approx 9.8\,\mathrm{m/s^2}\). In this study, all signals are standardized to \(\mathrm{m/s^2}\). Accordingly, data originally expressed in \textbf{g} are converted by multiplying by 9.8 to ensure consistency in quantitative comparisons.

\subsection{Feature Construction and Lightweight Classification Framework}

The final stage of the proposed method involves fusing the robust entropy metrics with a standard amplitude feature to construct a comprehensive measurement validity feature set, which is then fed into a lightweight machine learning model for automated signal discrimination.

\subsubsection{Feature Vector Construction}
The proposed MDE framework generates multiple entropy-based features from different domain perspectives, including time-domain, frequency-domain, and frequency-band representations. These features characterize signal complexity and distribution in a complementary manner. However, entropy-based indices mainly reflect distributional properties and are relatively insensitive to the absolute signal magnitude.

To compensate for this limitation while preserving computational efficiency, the Root Mean Square (RMS) value of the vibration signal is introduced as an auxiliary feature to represent the overall energy level.

Accordingly, the final multi-dimensional feature vector is constructed as:
\[
\mathbf{F} = [\mathrm{TAE}, \mathrm{ATAE}, \mathrm{SpAE\_1}, \mathrm{SpAE\_2}, \mathrm{FBEE\_1}, \mathrm{FBEE\_2}, \mathrm{RMS}].
\]

Here, TAE and ATAE are computed using a fixed bin width of $1/100$. Two variants of spectrum-amplitude entropy are defined: $\mathrm{SpAE\_1}$ uses a fixed bin width of $1/1000$, while $\mathrm{SpAE\_2}$ adopts an adaptive bin width proportional to the signal RMS (set as $\mathrm{RMS}/5$). Similarly, two frequency-band energy entropy measures are constructed: $\mathrm{FBEE\_1}$ partitions the spectrum into low- and high-frequency bands using a threshold of 500~Hz, whereas $\mathrm{FBEE\_2}$ employs uniform frequency bands with a width of 500~Hz.

This feature set provides a compact yet informative representation by jointly encoding time-domain amplitude distribution, spectral amplitude distribution, frequency-band energy distribution,and overall signal magnitude.

\subsubsection{Lightweight Anomaly Identification Model}

Three lightweight machine learning models are employed for the binary classification task (normal vs. erroneous vibration signals), including Support Vector Machine (SVM), Random Forest (RF), and Light Gradient Boosting Machine (LightGBM). These models are selected due to their strong generalization capability and relatively low computational overhead, making them suitable for practical deployment.

The overall classification procedure is as follows:
\begin{enumerate}
    \item \textbf{Data preprocessing:} Input signals are standardized to $\mathrm{m/s^2}$ and segmented.
    \item \textbf{Feature extraction:} The seven-dimensional feature vector $\mathbf{F}$ is extracted from each segment.
    \item \textbf{Model training:} The classify models are trained using labeled data.
    \item \textbf{Inference:} Features of a new signal segment are classified by the trained models to obtain the measurement validity status.
\end{enumerate}

\section{Experimental Design}
This section presents the experimental setup used to evaluate the effectiveness and robustness of the proposed MDE-based measurement validity assessment framework.
\subsection{Dataset Description}
The dataset employed in this study consists of real-world vibration measurements collected from wind turbine monitoring systems deployed across multiple wind farms. A detailed breakdown of the dataset is provided in Table~\ref{tab:origin_dataset}. In total, the dataset \textbf{covers 12 wind farms and 14 turbine units}, involving \textbf{2 turbine models (3.0 MW and 3.2 MW)}.

Vibration signals are acquired from \textbf{three key drivetrain components}, namely the main bearing, gearbox, and generator. The data are collected using \textbf{8 different sensor brands}, with heterogeneous measurement units (g and m/s$^2$), and \textbf{two sampling configurations}: 12800 Hz for 10.24 s and 25600 Hz for 5.12 s. To ensure consistency in feature extraction, all signals are converted into a unified unit of m/s$^2$.

The dataset spans \textbf{long-term monitoring records from 2019 to 2025}, covering diverse operating conditions and equipment configurations.The dataset comprises \textbf{57,643 vibration samples} in total. All samples were manually annotated according to predefined signal validity criteria based on domain knowledge. The diversity in sensor types, sampling configurations, turbine models, and operating conditions enables the dataset to effectively reflect realistic industrial scenarios.

For confidentiality reasons, specific project identifiers and turbine information have been anonymized. Due to proprietary agreements and data security requirements, the raw vibration datasets and the complete MDE feature datasets cannot be made publicly available. Upon reasonable request, a small batch of anonymized (de-identified) data—on the order of several hundred MDE feature samples—can be provided to support verification of the reported results.

\begin{table}[htbp]
  \centering
  \caption{The Information of Dataset}
  \label{tab:origin_dataset}
 
  \setlength{\tabcolsep}{3pt}
  \renewcommand{\arraystretch}{1.2}{
  \begin{tabular}{cccccccccc}
    \toprule 
    Model & \shortstack{Wind\\Farm} & Turbine & \shortstack{Sensor Brand} & Unit & Component & \shortstack{Sampling\\Freq. (Hz)} & \shortstack{Duration\\(s)} &\shortstack{Record\\Period} &\shortstack{Total\\Samples}\\
    \midrule 
    3.0 & A & a & Brand 1 &  g    & Main Bearing &  12800 &  10.24 &2022.11-2025.05 & 4646\\
    3.0 & B & b & Brand 2 & m/s$^{2}$ & Main Bearing &  12800 &  10.24 &2022.02-2025.05 &1400\\
    3.2 & C & c & Brand 3 & m/s$^{2}$  & Main Bearing &  12800 &  10.24 &2022.09-2025.07 &3683\\
    3.0 & D & d & Brand 4 &  g    & Main Bearing &  12800 &  10.24 &2019.04-2025.05 &1596\\
    3.2 & E & e & Brand 5 & m/s$^{2}$  & Main Bearing &  25600 &  5.12  &2024.02-2025.08 &2989\\
    \midrule 
    3.0 & F & f & Brand 6 & m/s$^{2}$  & Gearbox &  25600 &  5.12  &2022.09-2025.02 &1810\\
    3.2 & C & m & Brand 3 & m/s$^{2}$  & Gearbox &  12800/25600    &  10.24/5.12 &2022.09-2025.07 &3688/3683\\
    3.2 & G & g & Brand 7 &  g    & Gearbox &  12800 &  10.24 &2021.05-2025.02 &6130\\
    3.0 & H & h & Brand 8 &  g    & Gearbox &  12800 &  10.24 &2024.06-2025.05 &4684\\
    3.2 & I & i & Brand 8 &  g    & Gearbox &  12800 &  10.24 &2021.06-2025.05 &2018\\
    \midrule 
    3.0 & F & f & Brand 6 & m/s$^{2}$  & Generator &  25600 & 5.12  &2022.09-2025.02  &1810\\
    3.2 & J & j & Brand 1 &  g    & Generator &  25600 & 5.12  &2022.12-2025.05  &990\\
    3.2 & K & k & Brand 6 & m/s$^{2}$  & Generator &  25600 & 5.12  &2022.08-2023.01  &352\\
    3.2 & L & l & Brand 3 & m/s$^{2}$  & Generator &  25600 & 5.12  &2021.01-2025.02  &8558\\
    3.2 & L & n & Brand 3 & m/s$^{2}$  & Generator &  25600 & 5.12  &2022.09-2025.02  &9606\\
    \bottomrule 
  \end{tabular}
}
\end{table}

\subsection{Evaluation Metrics}
\label{evaluation metrics}
To quantitatively assess the performance of the measurement validity identification framework, four widely used evaluation metrics are employed: \textbf{Accuracy} (ACC), \textbf{True Detection Rate} (TDR), \textbf{False Detection Rate} (FDR), and \textbf{F1-score}. These metrics provide complementary perspectives on classify model performance and are defined as follows:

\begin{itemize}
    \item \textbf{Accuracy} (\textit{ACC}) measures the overall proportion of correctly classified samples among all test samples. It is expressed as:
    \[
        \text{Accuracy} = \frac{N_\text{TP} + N_\text{TN}}{N_\text{TP} + N_\text{TN} + N_\text{FP} + N_\text{FN}},
    \]
    where
    \begin{itemize}
        \item $N_\text{TP}$: number of true positives, i.e., erroneous vibration signals correctly identified as erroneous;
        \item $N_\text{TN}$: number of true negatives, i.e., normal operational data correctly identified as normal;
        \item $N_\text{FP}$: number of false positives, i.e., normal operational data incorrectly classified as erroneous;
        \item $N_\text{FN}$: number of false negatives, i.e., erroneous vibration signals incorrectly classified as normal.
    \end{itemize}

    \item \textbf{True Detection Rate} (\textit{TDR}), also known as \textit{recall}, reflects the proportion of correctly identified erroneous vibration signals among all actual erroneous vibration signals:
    \[
        \text{TDR} = \frac{N_\text{TP}}{N_\text{TP} + N_\text{FN}}.
    \]

    \item \textbf{False Detection Rate} (\textit{FDR}) represents the proportion of falsely flagged valid operational datas among all signals predicted as false:
    \[
        \text{FDR} = \frac{N_\text{FP}}{N_\text{TP} + N_\text{FP}}.
    \]

    \item \textbf{F1-score} is the harmonic mean of TDR and precision (where precision $= 1 - \text{FDR}$), providing a balanced measure when the class distribution is imbalanced:
    \[
        \text{F1-score} = 2 \cdot \frac{\text{Precision} \cdot \text{TDR}}{\text{Precision} + \text{TDR}}, \quad
        \text{Precision} = \frac{N_\text{TP}}{N_\text{TP} + N_\text{FP}}.
    \]
\end{itemize}

These indicators collectively enable comprehensive evaluation of the method’s ability to correctly identify false vibration signals while minimizing false alarms, which is critical for reliable data quality control in industrial wind turbine monitoring.

\subsection{Experimental Design and Validation}

All signals are preprocessed to ensure uniformity for feature extraction. As described in Section~\ref{MDE formula}, each signal is characterized by the proposed multi-dimensional feature set, including TAE, ATAE, SpAE\_1, SpAE\_2, FBEE\_1, FBEE\_2, and RMS, forming a 7-dimensional feature vector. The robustness verification of MDE under different sampling configurations (Section~\ref{Robustness of MDE under different sapling config}) demonstrates, through controlled simulations with varying noise levels, that signals collected with different sampling strategies (12800 Hz/10.24 s and 25600 Hz/5.12 s) exhibit only minor variations. Based on this, identical feature extraction is applied to all signals, irrespective of sampling frequency or duration.

A total of 57,643 vibration samples are used in this study. All samples are manually labeled by the authors with domain knowledge, ensuring the reliability of the ground truth. The dataset is randomly divided into training and testing subsets with a ratio of 3:1. The objective of this study is to evaluate the effectiveness of MDE in measurement validity identification under heterogeneous industrial conditions rather than to evaluate cross-farm generalization.

Three representative lightweight machine learning models—Support Vector Machine (SVM), Random Forest (RF), and LightGBM—are employed for comparative evaluation, while the proposed framework remains model-agnostic and can be extended to other classify models depending on specific application requirements.

All experiments were implemented in Python 3.10.13 on a workstation equipped with an Intel Core i5-12400F CPU and 32 GB RAM.

\section{Results}
\subsection{Overall Performance of Measurement Validity Identification}

To evaluate the effectiveness of the proposed MDE-based measurement validity representation, three lightweight machine learning models, including Support Vector Machine (SVM), Random Forest (RF), and LightGBM, are employed as evaluation tools. These models are widely adopted in industrial data analysis due to their computational efficiency and strong generalization capability.

The performance of each identification model is evaluated using the metrics defined in Section~\ref{evaluation metrics}. The robustness of the MDE features under different sampling configurations has been validated in Section~\ref{Robustness of MDE under different sapling config}.
\subsubsection{Hyperparameter Grid Search For Classification}
\label{hyperparameters search}
The hyperparameter optimization is performed using grid search on the training dataset. This process ensures that each model operates under near-optimal settings before final evaluation on the testing set.

For the SVM model, key hyperparameters such as the kernel function, regularization coefficient (C), kernel coefficient (gamma), and polynomial degree are tuned through grid search to balance the classification margin and model complexity. The polynomial degree parameter is only effective when the polynomial kernel is selected. The explored hyperparameter search space is summarized in Table~\ref{tab:svm-hyperparameters}.

\begin{table}[htbp]
\centering
\caption{Hyperparameter search space for the SVM classifier}
\label{tab:svm-hyperparameters}
\begin{tabular}{ll}
\toprule
Hyperparameter          & Candidate Values      \\
\midrule
Kernel & linear, rbf, poly, sigmoid \\
C      & 0.01, 0.1, 1, 10, 100 \\
Gamma  & scale, auto, 0.01, 0.1, 1 \\
Degree & 2, 3 \\
\bottomrule
\end{tabular}
\end{table}

For the Random Forest model, hyperparameters including the number of trees, maximum tree depth, and minimum samples split are tuned to enhance model stability while preventing overfitting. The explored hyperparameter search space is summarized in Table~\ref{tab:rf-hyperparameters}.

\begin{table}[htbp]
\centering
\caption{Hyperparameter search space for the Random Forest classifier}
\label{tab:rf-hyperparameters}
\begin{tabular}{ll}
\toprule
Hyperparameter          & Candidate Values      \\
\midrule
Number of Trees         & 50, 100, 200          \\
Max Tree Depth          & None, 10, 20, 30      \\
Min Samples Split       & 2, 5, 10              \\
\bottomrule
\end{tabular}
\end{table}
For the LightGBM model, key hyperparameters including the number of leaves, learning rate, maximum tree depth, and number of estimators (boosting iterations) are tuned to achieve high accuracy with efficient training. The explored hyperparameter search space is summarized in Table~\ref{tab:lightgbm-hyperparameters}. Note that a value of \textbf{-1} for maximum tree depth indicates \textbf{no depth restriction} (i.e., trees are allowed to grow until other stopping criteria, such as minimum child samples or purity, are met).

\begin{table}[htbp]
\centering
\caption{Hyperparameter search space for the LightGBM classifier}
\label{tab:lightgbm-hyperparameters}
\begin{tabular}{ll}
\toprule
Hyperparameter         & Candidate Values              \\
\midrule
Number of Leaves       & 15, 31, 63, 127                \\
Learning Rate          & 0.01, 0.05, 0.1, 0.2, 0.5, 1 \\
Max Tree Depth         & -1, 5, 7, 10                  \\
Number of Estimators   & 100, 200, 500, 1000           \\
\bottomrule
\end{tabular}
\end{table}

\subsubsection{Performance Comparison Between Models}

After performing grid search hyperparameter optimization as detailed in Section~\ref{hyperparameters search} (with the explored search spaces summarized in Tables~\ref{tab:svm-hyperparameters}, \ref{tab:rf-hyperparameters}, and \ref{tab:lightgbm-hyperparameters}), the best-performing hyperparameter configuration was selected for each identification model based on validation performance. The final models were then evaluated on the held-out testing dataset using the metrics defined in Section~\ref{evaluation metrics}: Accuracy, TDR, FDR, and F1-score.

Table~\ref{tab:classifier-performance} summarizes the performance of the three identification models on the testing set under their respective optimal Accuracy. Detailed results for more hyperparameter combinations evaluated during grid search are provided in Appendix~\ref{app:hyperparameters}.

\begin{table}[htbp]
\centering
\caption{The Performance of The Three Models On The Testing Set}
\label{tab:classifier-performance}
\begin{tabular}{l >{\raggedright}p{6cm} r r r r}
\toprule
Model & Best Hyperparameter & Accuracy & TDR   & FDR  & F1-score   \\
\midrule
SVM           & Kernel=rbf, C=100, Gamma=1, Degree=N/A                & 99.36\%  & 99.49\%  & 1.05\%  & 99.36\%  \\
\midrule
Random Forest & Number of Trees=200, Max Tree Depth=30, Min Sample Split=5 & 99.44\%  & 99.57\%  & 0.94\%  & 99.44\% \\
\midrule
LightGBM      & Number of Leaves=127, Learning Rate=0.05, Max Tree Depth=10, Number of Estimators=100  & 99.39\% & 99.41\% & 0.68\% & 99.39\%  \\
\bottomrule
\end{tabular}
\end{table}

Overall, all models demonstrate excellent identification capability for erroneous vibration measurements, with Accuracy, TDR, and F1-score exceeding 99\%, while the FDR remains close to zero. This indicates that the proposed MDE-based feature representation effectively captures the distinguishing characteristics of erroneous vibration measurements.

Among the evaluated models, Random Forest achieves slightly better performance, while all models consistently demonstrate the effectiveness of the proposed MDE representation.

Nevertheless, the consistently high scores across different identification models confirm that the feature representation itself plays a dominant role in the performance of the proposed approach, rather than relying solely on a specific model. These results demonstrate that the proposed method provides both high detection accuracy and strong model generalization, demonstrating its strong potential for practical deployment in industrial monitoring systems. Furthermore, the consistently high performance across different identification models suggests that the proposed MDE-based feature set provides a robust and model-independent representation for erroneous vibration signal identification.

\subsection{Visualization and Case Analysis}
To further illustrate the discriminative capability of the proposed MDE-based feature representation, both global visualization and statistical distribution analyses are conducted on the extracted feature space.

\begin{figure}[pos=htbp]
    \centering
    \setlength{\tabcolsep}{0pt}
    \begin{subfigure}{0.33\textwidth}
        \centering
        \includegraphics[width=\textwidth]{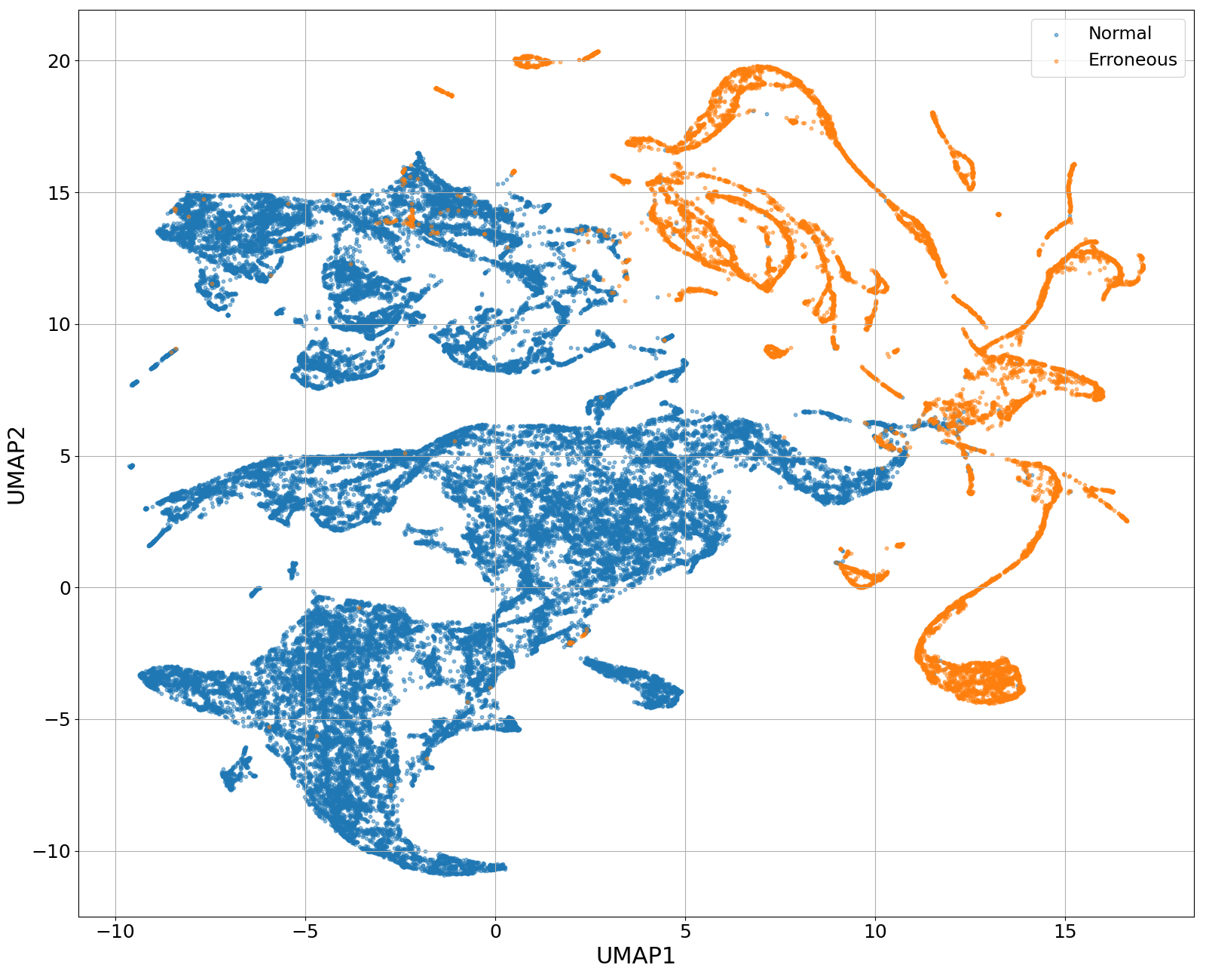}
        \caption{UMAP}
    \end{subfigure}\hspace{-2mm}
    \begin{subfigure}{0.33\textwidth}
        \centering
        \includegraphics[width=\textwidth]{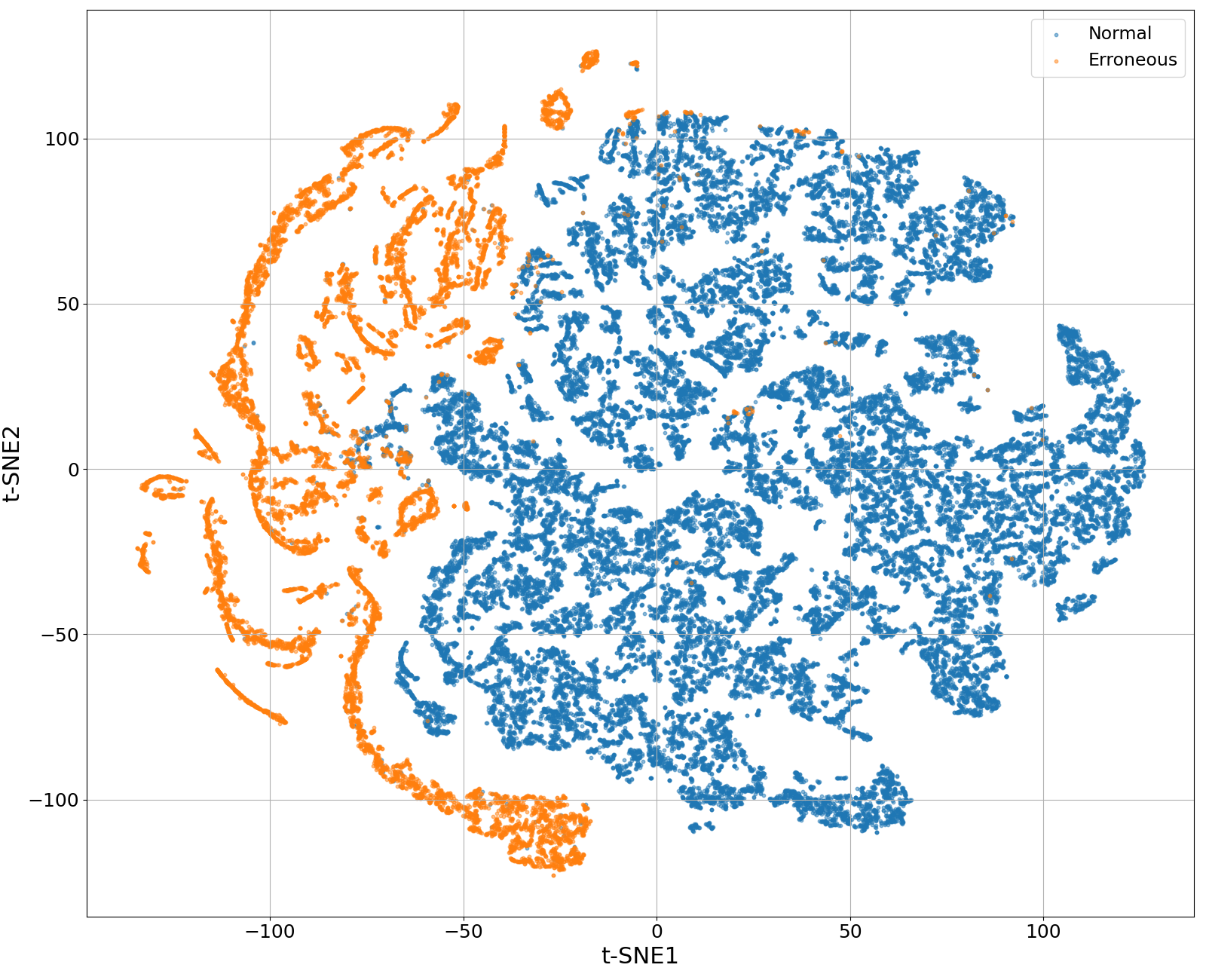}
        \caption{t-SNE}
    \end{subfigure}\hspace{-2mm}
    \begin{subfigure}{0.33\textwidth}
        \centering
        \includegraphics[width=\textwidth]{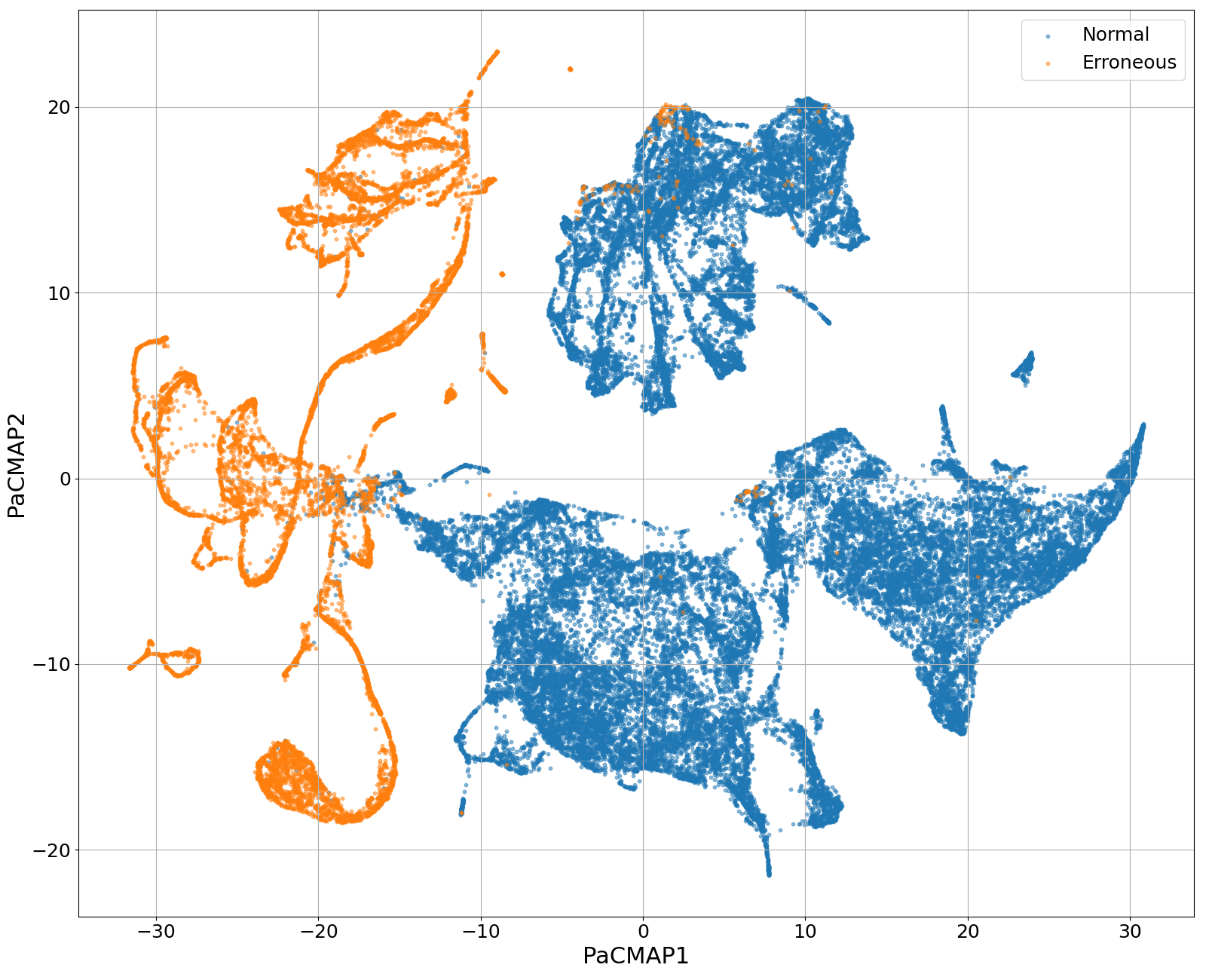}
        \caption{PaCMAP}
    \end{subfigure}
    
    \caption{Low-dimensional visualization of the MDE feature space using three manifold learning methods.}
    \label{fig:reduct_dimension_visual}
\end{figure}

First, low-dimensional projections of the feature vectors are obtained using three representative manifold learning techniques, namely UMAP, t-SNE, and PaCMAP \cite{McInnes2018,VanDerMaaten2008,Wang2021PaCMAP}, as shown in Figure~\ref{fig:reduct_dimension_visual}. Each point represents a vibration sample, where blue denotes normal signals and orange denotes erroneous signals. Despite the heterogeneity of the dataset in terms of wind farms, turbine units, sampling configurations, etc., all three methods consistently reveal that the majority of normal and erroneous samples are clearly separable, with only a small portion exhibiting overlap. This indicates that the proposed MDE feature set provides strong intrinsic separability in the feature space.

\begin{figure}[pos=htbp]
\centering
\includegraphics[width=1.05\textwidth]{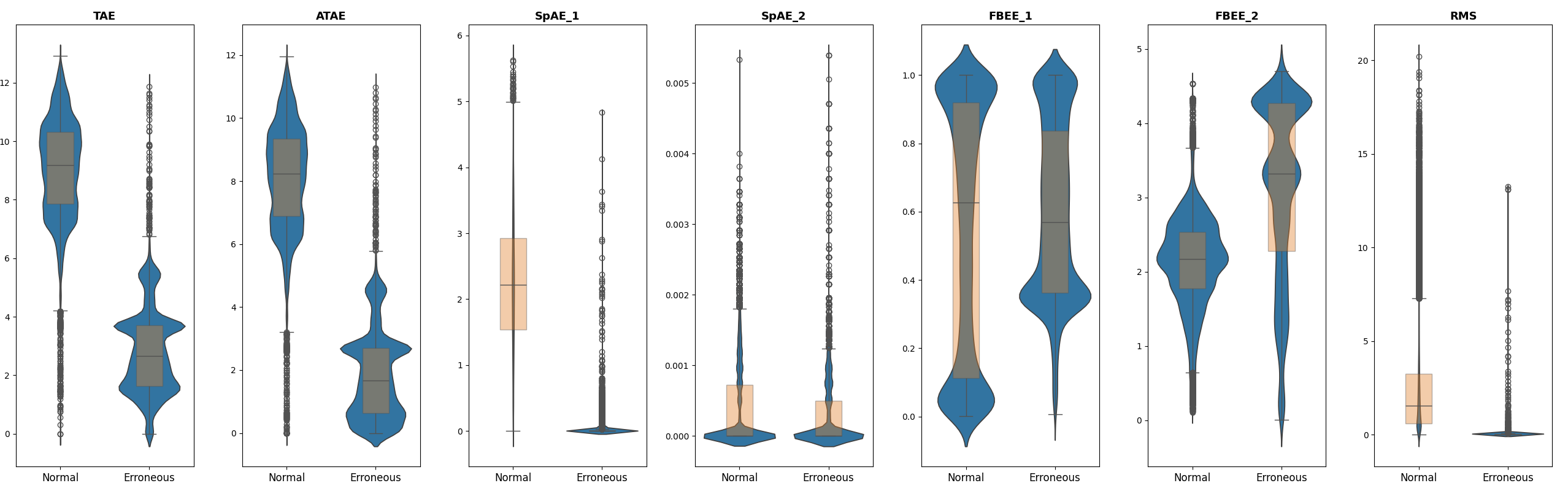}
\caption{Statistical distribution of MDE features for normal and erroneous vibration signals using boxplots and violin plots.}
\label{fig:box violin plot}
\end{figure}

To further investigate the distributional differences of individual features, statistical visualizations are provided in Figure~\ref{fig:box violin plot}, where boxplots and violin plots are jointly used to depict the feature distributions for normal and erroneous signals. The results show that time-domain entropy features (TAE and ATAE) exhibit significantly higher values for erroneous signals compared to normal signals, indicating more concentrated amplitude distributions in abnormal measurements. 

Compared with other entropy-based features, TAE and ATAE present clear separation between normal and erroneous signals. Meanwhile, features such as SpAE\_2 show relatively limited discriminative power, as evidenced by their highly overlapping distributions.

Overall, these visualization and statistical analysis results confirm that the proposed feature representation not only achieves high identification performance but also provides physically interpretable distinctions between normal and erroneous vibration signals.

\subsection{Feature Importance and Contribution Analysis}
\label{feature importance and contribution analysis}
Feature importance analysis is conducted to quantify the relative contributions of individual features and to assess their robustness across different models. For both Random Forest and LightGBM, importance scores are normalized to sum to 1 (analogous to Gini importance) and averaged over 10 independent runs with different random seeds to account for stochastic variability (detailed per-run values are provided in Appendix~\ref{app:feature-importance}).

\begin{figure}[pos=htbp]
\centering
\begin{subfigure}[b]{0.49\textwidth}
\centering
\includegraphics[width=\textwidth]{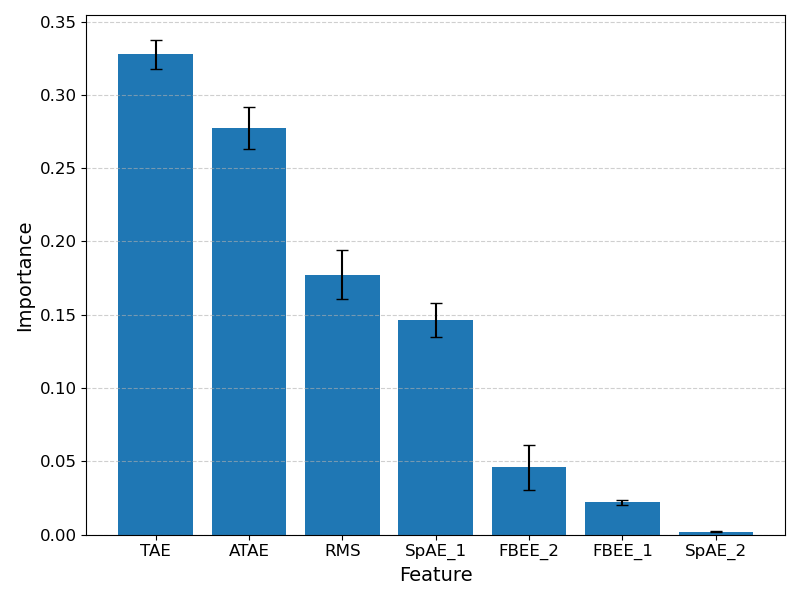} 
\caption{Random Forest: mean Gini importance}
\label{fig:rf-importance}
\end{subfigure}
\hfill
\begin{subfigure}[b]{0.49\textwidth}
\centering
\includegraphics[width=\textwidth]{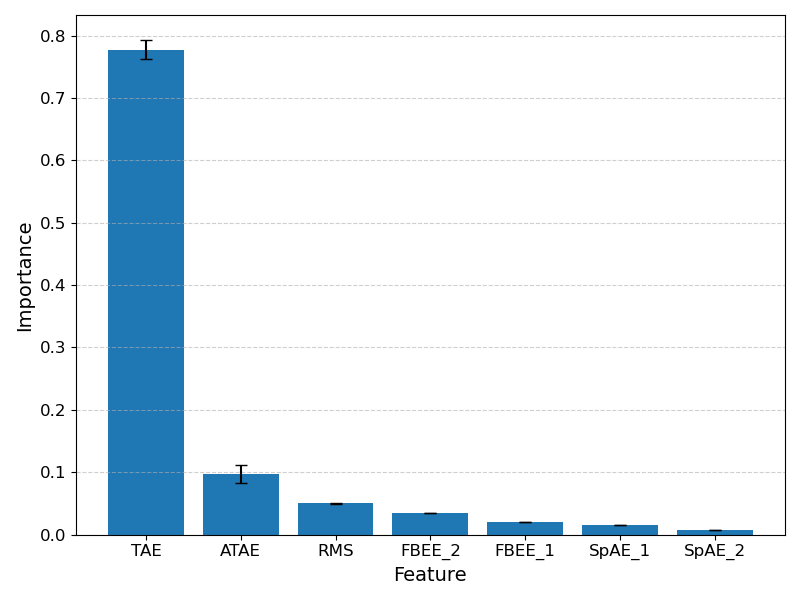} 
\caption{LightGBM: mean normalized split importance}
\label{fig:lightgbm-importance}
\end{subfigure}
\caption{Comparison of feature importance distributions between Random Forest and LightGBM. Error bars indicate one standard deviation across 10 runs. The figure highlights the distinct allocation patterns of feature contributions, where Random Forest exhibits a relatively balanced distribution, while LightGBM is strongly dominated by TAE.}
\label{fig:feature-importance}
\end{figure}


In the Random Forest model (Figure~\ref{fig:rf-importance}), feature importance is relatively evenly distributed among the top contributors. As summarized in Table~\ref{tab:feature-importance-summary}, TAE (0.3277 ± 0.0098), ATAE (0.2775 ± 0.0144), and RMS (0.1774 ± 0.0166) rank as the three most influential features. No single feature dominates excessively, reflecting the bagging mechanism's tendency to average contributions across multiple decision trees.


In contrast, LightGBM (Figure~\ref{fig:lightgbm-importance}) exhibits a markedly different pattern, the detailed per-run results in Table~\ref{tab:app-lightgbm-importance}. TAE overwhelmingly dominates with an importance of 0.7776 ± 0.0158, consistently exceeding 50\% across all runs and reaching up to 0.9042. The remaining features contribute only marginally, with ATAE and RMS ranking second and third, respectively. This concentration is characteristic of boosting algorithms, where early discriminative splits are repeatedly reinforced during training.

\begin{table}[htbp]
    \centering
    \caption{Summary of average feature importance (mean ± std) across 10 runs for Random Forest and LightGBM}
    \label{tab:feature-importance-summary}
    \begin{tabular}{l|c|c}
    \toprule
        Feature     &  Random Forest(mean ± std) & LightGBM(mean ± std)\\
    \midrule
        TAE         & 0.3277 ± 0.0098 & 0.7776 ± 0.0158 \\
        ATAE        & 0.2775 ± 0.0144 & 0.0968 ± 0.0142 \\
        RMS         & 0.1774 ± 0.0166 & 0.0498 ± $4.242\times10^{-4}$ \\    
        SpAE\_1     & 0.1465 ± 0.0116 & 0.0153 ± $2.741\times10^{-6}$ \\
        FBEE\_2     & 0.0460 ± 0.0154 & 0.0344 ± $3.82\times10^{-6}$ \\
        FBEE\_1     & 0.0219 ± 0.0017 & 0.0195 ± $6.846\times10^{-6}$ \\
        SpAE\_2     & 0.0020 ± 0.0002 & 0.0066 ± $3.637\times10^{-7}$ \\
    \bottomrule
    \end{tabular}

\end{table}

Despite these distributional differences—which arise from the fundamental distinctions between bagging (RF) and boosting (LightGBM) mechanisms—the two models show strong agreement on the most critical feature: TAE and  ATAE consistently ranks top 2 in both cases, highlighting its dominant role in capturing time-domain structural irregularities in erroneous vibration signals. The RMS feature consistently ranks third in RF and occupies the third position in the majority of LightGBM cases, highlighting its complementary contribution to overall signal energy. Other entropy-based indicators provide moderate, supportive contributions, whereas SpAE\_2 exhibits negligible influence across all runs.

These findings further validate the effectiveness and robustness of the proposed MDE-based feature representation. The consistent importance ranking across different ensemble strategies (bagging vs. boosting) and multiple random initializations indicates that the extracted features capture intrinsic signal characteristics rather than model-specific artifacts, reinforcing the robustness of the proposed approach under heterogeneous industrial measurement conditions.

\section{Discussion}

This study introduces a class of entropy-based indicators termed Multi-Dimensional Entropy (MDE) for vibration data quality control. The main motivation of MDE is to quantify vibration signals from multiple perspectives, including time-domain and frequency-domain representations, through information entropy measurements. Compared with traditional single-domain descriptors, this multi-dimensional framework provides a more comprehensive characterization of signal distribution patterns. Such capability is particularly important in industrial monitoring environments, where vibration data are often affected by sensor anomalies, operational interruptions, or measurement disturbances. From a practical perspective, reliable reliable data quality control serves as a fundamental prerequisite for automated monitoring and diagnostic systems. As industrial monitoring systems increasingly integrate artificial intelligence algorithms, separating erroneous vibration signals from valid operational data becomes an essential preprocessing step. In this work, MDE indicators are combined with lightweight machine learning models to perform vibration data quality control, enabling the separation of erroneous vibration signals from normal operational signals.

The experimental results demonstrate that the proposed entropy features, together with the RMS representing signal energy, can effectively characterize the distribution differences between normal and erroneous vibration signals. Hyperparameter search experiments were conducted for three lightweight machine learning models under consistent experimental settings. For each model, the optimal configuration achieved accuracy and F1-score values exceeding 99\%, with only minor variations across models. These results are derived from a large-scale real-world dataset characterized by substantial heterogeneity. The dataset covers 12 wind farms and 14 turbine units, involving two turbine models (3.0 MW and 3.2 MW), three key drivetrain components, and data collected from eight different sensor brands. In addition, the dataset includes multiple sampling configurations (12800 Hz/10.24 s and 25600 Hz/5.12 s) and spans long-term monitoring periods from 2019 to 2025. Such diversity in sensor types, acquisition settings, turbine models, and operational conditions introduces significant variability, which closely reflects practical industrial environments. All samples were manually annotated according to predefined signal validity criteria based on domain knowledge. Consequently, some samples identified as erroneous by the model may correspond to labeling inaccuracies. In such cases, the discriminative capability of the proposed MDE indicators may be even stronger than reflected by the reported results.

To further investigate the contribution of individual features, feature importance analysis was conducted for the Random Forest and LightGBM models. The reported importance values follow the format of mean ± standard deviation, obtained from ten independent runs with different random seeds. To ensure comparability, identical random seeds were used across models within each run.

The results show that time-domain entropy features play a dominant role in distinguishing normal and erroneous vibration signals. In the Random Forest model, the two time-domain entropy features contributed approximately $0.6052 \pm 0.0134$ of the total importance. When combined with RMS, the cumulative importance of these three features increased to $0.7826 \pm 0.02$, indicating that these features capture the majority of discriminative information required for validity identification. A similar but even more pronounced pattern was observed in the LightGBM model. Among all features, the indicator TAE consistently showed a leading contribution, reaching $0.7776 \pm 0.0158$ on average. When both time-domain entropy features were considered together, their cumulative importance increased to $0.8744 \pm 0.0004$, and further increased to $0.9242 \pm 2.226 \times 10^{-5}$ when RMS is included. These observations indicate that the combination of time-domain entropy and energy-related features forms the primary basis for distinguishing erroneous signals, while the remaining entropy indicators mainly provide complementary information.

The dominance of time-domain entropy features is also consistent with the signal characteristics observed in the waveform examples presented in Section~\ref{feature importance and contribution analysis} and Appendix~\ref{app:feature-importance}. Erroneous vibration signals typically exhibit relatively concentrated amplitude distributions, whereas normal vibration signals tend to show broader amplitude variations. As a result, entropy measures derived from the time domain are particularly sensitive to such differences in amplitude distribution. Notably, it was initially expected that entropy features from different domains would contribute more evenly to the validity identification task. However, the experimental results indicate that, for tree-based models such as Random Forest and LightGBM, time-domain entropy features provide significantly stronger discriminative power. This finding suggests that the temporal structure of vibration signals may contain more direct information about abnormal measurement states than previously assumed. The underlying reasons for this phenomenon, as well as its implications for feature design in vibration analysis, deserve further investigation in future work.

Another important observation is the \textbf{robustness of the proposed MDE framework}. The experimental dataset spans different wind farms, turbine types, and operational periods. Despite these variations, the proposed method consistently achieved identification performance above 99\% across all three machine learning models. Furthermore, the time-domain entropy features demonstrate strong stability in importance rankings across repeated experiments, indicating that their discriminative capability is not sensitive to variations in model type or hyperparameter settings. These results suggest that the MDE indicators possess strong robustness under heterogeneous industrial measurement conditions.

Beyond the specific application investigated in this study, the proposed MDE framework may also have broader practical implications. In industrial monitoring systems, erroneous vibration signals may arise from sensor faults, equipment shutdown conditions, or abnormal measurement states. Similar issues may occur in other industrial domains such as chemical processing, marine engineering, aerospace systems, and rail transportation. As industrial diagnostic platforms gradually evolve toward fully automated analysis systems, \textbf{reliable data quality control will become an essential prerequisite for trustworthy automated monitoring systems.} Without separating erroneous signals from valid operational data, diagnostic algorithms may produce misleading results. In this context, the MDE framework can serve as an effective preprocessing step for vibration-based monitoring systems. Due to its computational simplicity and compatibility with lightweight machine learning models, the proposed method can be conveniently integrated into existing diagnostic software architectures. Moreover, MDE have the potential  ablity to be deployed in edge environments in the future, such as wind farm control centers and embedded monitoring systems. The low hardware dependency further enhances its practical applicability, enabling potential applicability to other industrial monitoring scenarios.

Future work will focus on extending the current method in several directions. First, additional validation models will be explored beyond the three models investigated in this study. Second, the robustness and contribution of individual MDE features will be further examined through systematic ablation studies, providing deeper insights into the mechanism and stability of the proposed entropy-based indicators.

\section{Conclusion}

This study investigates the problem of vibration data quality control in industrial monitoring systems, with particular focus on distinguishing erroneous vibration signals from valid operational data. Reliable data quality control is a fundamental prerequisite for the development and deployment of automated monitoring systems in modern industrial environments.

To address this issue, a class of entropy-based indicators termed \textbf{Multi-Dimensional Entropy (MDE)} is proposed. The MDE framework characterizes vibration signals from multiple perspectives, including time-domain amplitude distribution, spectral amplitude distribution, and frequency band energy distribution. By integrating these entropy descriptors with the RMS feature, a compact feature representation of vibration signals was constructed. The resulting feature vectors were then combined with lightweight machine learning models to perform vibration data quality control.

Experiments were conducted on real-world monitoring data collected from multiple wind farms, comprising over 57,000 labeled samples. The experimental results demonstrate that the proposed approach achieves classification performance exceeding 99\% in terms of both accuracy and F1-score across multiple machine learning models. Feature importance analysis further reveals that time-domain entropy, together with RMS features, capture the majority of discriminative information required to distinguish erroneous signals from normal vibration signals.

Overall, the proposed MDE framework provides an effective and computationally efficient solution for vibration data quality control. Owing to its simplicity, interpretability, and compatibility with lightweight machine learning models, the method can be readily integrated into industrial monitoring and diagnostic systems, and is well-suited for deployment in resource-constrained environments.

Future work will focus on further investigating the robustness of the MDE framework through ablation studies and exploring its applicability to additional machine learning models and industrial monitoring scenarios.

\section*{Acknowledgements}
The author would like to thank Mhamed Djellali for his invaluable assistance in refining the manuscript's English expression and providing insightful suggestions.

\clearpage

\appendix

\section{Hyperparameter Search Details}
\label{app:hyperparameters}
This appendix provides selected results of the grid search hyperparameter tuning performed directly on the training set for each model. The best-performing configuration for each model was selected based on these training results and then evaluated on the held-out testing set (as reported in Section 5.1 and Table~\ref{tab:classifier-performance}).

Due to space limitations, only a subset of the evaluated combinations is presented below, along with their corresponding training performance metrics.

Complete grid search results, including all models, all hyperparameter combinations and corresponding metrics, are available in the supplementary material or upon request.

\subsection{SVM Classifier}
The grid search encompassed 4 kernel functions, 5 regularization coefficients (C), 5 kernel coefficients (gamma), and 2 polynomial degrees (degree effective solely for the polynomial kernel), resulting in up to 125 evaluated combinations. Table~\ref{tab:app-svm-top15-detailed} lists the top 15 configurations sorted by descending F1-score on the testing set, together with their detailed performance metrics.

\begin{table}[htbp]
\centering
\caption{Top-15 hyperparameter configurations for SVM with detailed performance metrics on the testing set}
\label{tab:app-svm-top15-detailed}
\small
\begin{tabular}{lccccrrrrrrrr}
\toprule
Rank & Kernel & C   & Gamma & Degree &$N_\text{TP}$&$N_\text{FN}$&$N_\text{FP}$&$N_\text{TN}$& ACC & TDR     & FDR     & F1-score \\
\midrule
1   & rbf    & 100  & 1    & None & 10766 & 55 & 37 & 3474 & 99.36\% & 99.49\% & 1.05\% & 99.36\% \\
2   & rbf    & 10   & 1    & None & 10764 & 57 & 38 & 3473 & 99.34\% & 99.47\% & 1.08\% & 99.34\% \\
3   & rbf    & 1    & 1    & None & 10769 & 52 & 48 & 3463 & 99.30\% & 99.52\% & 1.37\% & 99.30\% \\
4   & rbf    & 100  & auto & None & 10762 & 59 & 45 & 3466 & 99.27\% & 99.45\% & 1.28\% & 99.27\% \\
5   & rbf    & 100  & 0.1  & None & 10762 & 59 & 50 & 3461 & 99.24\% & 99.45\% & 1.42\% & 99.24\% \\
6   & rbf    & 10   & auto & None & 10767 & 64 & 56 & 3455 & 99.23\% & 99.50\% & 1.60\% & 99.23\% \\
7   & rbf    & 10   & 0.1  & None & 10764 & 57 & 63 & 3448 & 99.16\% & 99.47\% & 1.79\% & 99.16\% \\
8   & rbf    & 0.1  & 1    & None & 10769 & 52 & 68 & 3443 & 99.16\% & 99.52\% & 1.94\% & 99.16\% \\
9   & rbf    & 1    & auto & None & 10757 & 64 & 65 & 3446 & 99.10\% & 99.41\% & 1.85\% & 99.10\% \\
10  & rbf    & 1    & 0.1  & None & 10748 & 73 & 64 & 3447 & 99.04\% & 99.33\% & 1.82\% & 99.04\% \\
11  & rbf    & 0.01 & 1    & None & 10772 & 49 & 91 & 3420 & 99.02\% & 99.55\% & 2.60\% & 99.02\% \\
12  & rbf    & 0.1  & 0.1  & None & 10748 & 73 & 69 & 3442 & 99.00\% & 99.33\% & 1.97\% & 99.00\% \\
13  & rbf    & 100  & auto & None & 10728 & 93 & 55 & 3456 & 98.97\% & 99.14\% & 1.57\% & 98.97\% \\
14  & poly   & 1    & auto & 3    & 10716 & 105& 45 & 3466 & 98.95\% & 99.03\% & 1.28\% & 98.97\% \\
15  & poly   & 10   & 0.1  & 2    & 10726 & 95 & 56 & 3455 & 98.94\% & 99.12\% & 1.60\% & 98.95\% \\
\bottomrule
\end{tabular}
\end{table}

\subsection{Random Forest Classifier}

The grid search explored 3 values for number of trees, 4 values for maximum tree depth, and 3 values for minimum samples split, resulting in 3 × 4 × 3 = 36 combinations (as summarized in Table~\ref{tab:rf-hyperparameters}). Table~\ref{tab:app-rf-top10-detailed} below lists the top-10 configurations ranked by descending F1-score on the testing set, together with their detailed performance metrics.

\begin{table}[htbp]
\centering
\caption{Top-10 hyperparameter configurations for Random Forest with detailed performance metrics on the testing set}
\label{tab:app-rf-top10-detailed}
\small
\begin{tabular}{lcccrrrrrrrr}
\toprule
Rank&\makecell{Number\\of\\Trees}&\makecell{Max\\Tree\\Depth}&\makecell{Min\\Samples\\Split}&$N_\text{TP}$&$N_\text{FN}$&$N_\text{FP}$&$N_\text{TN}$& ACC & TDR   & FDR   & F1-score \\
\midrule
1    & 200          & 30        & 5             & 10774 & 47 & 33 & 3478 & 99.44\%  & 99.57\% & 0.94\%  & 99.44\% \\
1    & 200          & 20        & 2             & 10774 & 47 & 33 & 3478 & 99.44\%  & 99.57\% & 0.94\%  & 99.44\% \\
1    & 200          & None      & 5             & 10774 & 47 & 33 & 3478 & 99.44\%  & 99.57\% & 0.94\%  & 99.44\% \\
4    & 50           & 10        & 2             & 10778 & 43 & 37 & 3474 & 99.44\%  & 98.60\% & 1.05\%  & 99.44\% \\
5    & 100          & 20        & 10            & 10774 & 47 & 34 & 3477 & 99.43\%  & 99.57\% & 0.97\%  & 99.44\% \\
6    & 200          & 20        & 5             & 10773 & 48 & 34 & 3477 & 99.43\%  & 99.56\% & 0.97\%  & 99.43\% \\
6    & 100          & 30        & 5             & 10773 & 48 & 34 & 3477 & 99.43\%  & 99.56\% & 0.97\%  & 99.43\% \\
6    & 100          & None      & 5             & 10773 & 48 & 34 & 3477 & 99.43\%  & 99.56\% & 0.97\%  & 99.43\% \\
9    & 50           & 20        & 10            & 10774 & 47 & 35 & 3476 & 99.43\%  & 99.57\% & 1.00\%  & 99.43\% \\
10   & 100          & 20        & 5             & 10775 & 46 & 36 & 3475 & 99.43\%  & 99.58\% & 1.03\%  & 99.43\% \\
\bottomrule
\end{tabular}
\end{table}

\subsection{LightGBM Classifier}
The grid search explored 4 values for number of leaves, 6 values for learning rate, 4 values for maximum tree depth, and 4 values for number of estimators, resulting in 4 × 6 × 4 × 4 = 384 combinations (as summarized in Table~\ref{tab:lightgbm-hyperparameters}).Table~\ref{tab:app-lightgbm-top10-detailed} lists the top-10 performing configurations ranked by descending F1-score on the testing set, together with their detailed performance metrics.

\begin{table}[htbp]
\centering
\caption{Top-10 hyperparameter configurations for LightGBM with detailed performance metrics on the testing set}
\label{tab:app-lightgbm-top10-detailed}
\small
\begin{tabular}{ccccrrrrrrrr}
\toprule
\makecell{Number\\of\\Leaves}&\makecell{Learning\\Rate}&\makecell{Max\\Tree\\Depth}&\makecell{Number\\of\\Estimators}&$N_\text{TP}$&$N_\text{FN}$&$N_\text{FP}$&$N_\text{TN}$& ACC & TDR   & FDR   & F1-score \\
\midrule
127 & 0.05 & 10  & 100 & 10757 & 64 & 24 & 3478 & 99.39\%  & 99.41\% & 0.68\%  & 99.39\% \\
127 & 0.1  & 7   & 200 & 10760 & 61 & 27 & 3484 & 99.39\%  & 99.44\% & 0.77\%  & 99.39\% \\
31  & 0.05 & -1  & 500 & 10764 & 57 & 31 & 3480 & 99.39\%  & 99.47\% & 0.88\%  & 99.39\% \\
31  & 0.05 & -1  & 200 & 10757 & 64 & 25 & 3486 & 99.38\%  & 99.41\% & 0.71\%  & 99.38\% \\
127 & 0.05 & -1  & 100 & 10760 & 61 & 28 & 3483 & 99.38\%  & 99.44\% & 0.80\%  & 99.38\% \\
63  & 0.05 & -1  & 100 & 10760 & 61 & 28 & 3483 & 99.38\%  & 99.44\% & 0.80\%  & 99.38\% \\
63  & 0.1  & 10  & 200 & 10762 & 59 & 30 & 3481 & 99.38\%  & 99.45\% & 0.85\%  & 99.38\% \\
63  & 0.2  & 7   & 200 & 10762 & 59 & 30 & 3481 & 99.38\%  & 99.45\% & 0.85\%  & 99.38\% \\
31  & 0.05 & 7   & 500 & 10762 & 59 & 30 & 3481 & 99.38\%  & 99.45\% & 0.85\%  & 99.38\% \\
31  & 0.05 & 10  & 500 & 10765 & 56 & 33 & 3478 & 99.38\%  & 99.48\% & 0.94\%  & 99.38\% \\
\bottomrule
\end{tabular}
\end{table}

\section{Feature Importance Details}
\label{app:feature-importance}

This appendix provides the detailed feature importance results for the Random Forest and LightGBM models, as summarized in Figure~\ref{fig:feature-importance} (Section 5.3). For Random Forest, importance scores were calculated using mean decrease in impurity (Gini importance). For LightGBM, normalized split importance was adopted for consistency in comparison. To evaluate robustness and account for stochastic variability, each model was trained 10 times using different random seeds. The tables below report the per-run importance values, together with the average and standard deviation across the 10 independent runs.

\subsection{Random Forest Feature Importance}

Table~\ref{tab:app-rf-importance} presents the per-run Gini importance scores, along with their mean and standard deviation across 10 independent runs. The results exhibit high consistency: the ranking of features by importance remains identical across all runs. This strong stability allows the features to be ordered by descending average importance in the table.

\begin{table}[htbp]
\centering
\caption{Feature importance values from Random Forest across 10 independent runs with different random seeds}
\label{tab:app-rf-importance}
\footnotesize
\begin{tabular}{l *{10}{c} cc}
\toprule
Feature&Seed 1&Seed 2&Seed 3&Seed 4&Seed 5&Seed 6&Seed 7&Seed 8&Seed 9&Seed 10&Average&Std. Dev. \\
\midrule
TAE      & 0.3232 & 0.3550 & 0.3216 & 0.3258 & 0.3242 & 0.3252 & 0.3282 & 0.3224 & 0.3276 & 0.3234 & 0.3277 & 0.0098 \\
ATAE     & 0.2951 & 0.2655 & 0.2940 & 0.2788 & 0.2645 & 0.2941 & 0.2616 & 0.2822 & 0.2576 & 0.2818 & 0.2775 & 0.0144 \\
RMS      & 0.1634 & 0.1948 & 0.1550 & 0.1925 & 0.1946 & 0.1643 & 0.1662 & 0.1749 & 0.1674 & 0.2010 & 0.1774 & 0.0166 \\
SpAE\_1  & 0.1346 & 0.1351 & 0.1483 & 0.1525 & 0.1666 & 0.1396 & 0.1565 & 0.1395 & 0.1584 & 0.1339 & 0.1465 & 0.0116 \\
FBEE\_2  & 0.0571 & 0.0280 & 0.0545 & 0.0257 & 0.0274 & 0.0509 & 0.0626 & 0.0536 & 0.0652 & 0.0348 & 0.0460 & 0.0154 \\
FBEE\_1  & 0.0236 & 0.0186 & 0.0232 & 0.0216 & 0.0202 & 0.0231 & 0.0221 & 0.0242 & 0.0209 & 0.0218 & 0.0219 & 0.0017 \\
SpAE\_2  & 0.0031 & 0.0031 & 0.0034 & 0.0031 & 0.0027 & 0.0027 & 0.0028 & 0.0031 & 0.0029 & 0.0032 & 0.0020 & 0.0002 \\

\bottomrule
\end{tabular}
\end{table}

\subsection{LightGBM Feature Importance}

Table~\ref{tab:app-lightgbm-importance} reports the per-run normalized split importance scores, together with their mean and standard deviation across 10 independent runs. In contrast to Random Forest, the feature importance rankings in LightGBM show greater variability across runs. While \textbf{time\_entropy\_1} consistently ranks as the most important feature in every run, the relative ordering of the remaining features fluctuates. Therefore, the table presents the features in the same order as they appear in the dataset.
\begin{table}[htbp]
\centering
\caption{Feature importance values from LightGBM across 10 independent runs with different random seeds}
\label{tab:app-lightgbm-importance}
\footnotesize
\begin{tabular}{l *{10}{c} cc}
\toprule
Feature&Seed 1&Seed 2&Seed 3&Seed 4&Seed 5&Seed 6&Seed 7&Seed 8&Seed 9&Seed 10&Average&Std. Dev. \\
\midrule
TAE      & 0.6378 & 0.9042 & 0.8334 & 0.8385 & 0.5605 & 0.8726 & 0.6313 & 0.8796 & 0.8799 & 0.7385 & 0.7776 & 0.0158 \\
ATAE     & 0.2266 & 0.0060 & 0.0618 & 0.0451 & 0.3091 & 0.0045 & 0.2420 & 0.0034 & 0.0068 & 0.1179 & 0.0968 & 0.0142 \\
SpAE\_1  & 0.0143 & 0.0167 & 0.0147 & 0.0168 & 0.0165 & 0.0121 & 0.0145 & 0.0173 & 0.0163 & 0.0139 & 0.0153 & $2.741 \times 10^{-6}$ \\
SpAE\_2  & 0.0069 & 0.0072 & 0.0071 & 0.0064 & 0.0064 & 0.0053 & 0.0062 & 0.0066 & 0.0073 & 0.0066 & 0.0066 & $3.637 \times 10^{-7}$ \\
FBEE\_1  & 0.0207 & 0.0185 & 0.0205 & 0.0194 & 0.0187 & 0.0139 & 0.0199 & 0.0244 & 0.0208 & 0.0185 & 0.0195 & $6.846 \times 10^{-6}$ \\
FBEE\_2  & 0.0332 & 0.0367 & 0.0352 & 0.0342 & 0.0375 & 0.0329 & 0.0356 & 0.0314 & 0.0347 & 0.0331 & 0.0344 & $3.82 \times 10^{-6}$ \\
RMS      & 0.0618 & 0.0107 & 0.0829 & 0.0395 & 0.0513 & 0.0587 & 0.0504 & 0.0373 & 0.0343 & 0.0714 & 0.0498 & $4.242 \times 10^{-4}$ \\
\bottomrule
\end{tabular}
\end{table}

\printcredits

\bibliographystyle{cas-model2-names}

\bibliography{cas-refs}

\end{document}